\documentclass[authoryear,review,11pt,a4paper]{elsarticle}

\usepackage{setspace}
\usepackage{fullpage}

\usepackage{color}

\usepackage{amsmath}

\usepackage{amsfonts}
\usepackage{amssymb}
\usepackage{amsthm}

\usepackage{amscd}

\usepackage{bbm}

\usepackage[ruled]{algorithm2e}

\usepackage{graphicx}

\usepackage{lscape}

\usepackage{booktabs}
\usepackage{multirow}
\usepackage{float}
\usepackage{caption}
\usepackage{subcaption}

\usepackage{lineno}

\usepackage{hyperref}

\biboptions{authoryear}

\newdefinition{rmk}{Remark}
\newproof{pf}{Proof}

\journal{ }

\begin{document}

\begin{frontmatter}



\title{From aggressive to conservative early stopping in Bayesian group sequential designs}

\author[TGIUK,ICTU]{Zhangyi He\corref{cor1}\fnref{fn3}}
\ead{zhe1@georgeinstitute.org.uk}

\author[BrisMath]{Feng Yu}

\author[ICTU]{Suzie Cro}

\author[TGIAUS]{Laurent Billot}

\cortext[cor1]{Corresponding author.}
\fntext[fn3]{Affiliation at time of study}

\address[TGIUK]{The George Institute for Global Health, Imperial College London, London W12 7RZ, United Kingdom}
\address[ICTU]{Imperial Clinical Trials Unit, Imperial College London, London W12 7RH, United Kingdom}
\address[BrisMath]{School of Mathematics, University of Bristol, Bristol BS8 1QU, United Kingdom}
\address[TGIAUS]{The George Institute for Global Health, University of New South Wales, Sydney 2042, Australia}

\begin{abstract}
Group sequential designs (GSDs) are widely used in confirmatory trials to allow interim monitoring while preserving control of the type I error rate. In the frequentist framework, O'Brien-Fleming-type stopping boundaries dominate practice because they impose highly conservative early stopping while allowing more liberal decisions as information accumulates. Bayesian GSDs, in contrast, are most often implemented using fixed posterior probability thresholds applied uniformly at all analyses. While such designs can be calibrated to control the overall type I error rate, they do not penalise early analyses and can therefore lead to substantially more aggressive early stopping. Such behaviour can risk premature conclusions and inflation of treatment effect estimates, raising concerns for confirmatory trials. We introduce two practically implementable refinements that restore conservative early stopping in Bayesian GSDs. The first introduces a two-phase structure for posterior probability thresholds, applying more stringent criteria in the early phase of the trial and relaxing them later to preserve power. The second replaces posterior probability monitoring at interim looks with predictive probability criteria, which naturally account for uncertainty in future data and therefore suppress premature stopping. Both strategies require only one additional tuning parameter and can be efficiently calibrated. In the HYPRESS setting, both approaches achieve higher power than the conventional Bayesian design while producing alpha-spending profiles closely aligned with O'Brien-Fleming-type behaviour at early looks. These refinements provide a principled and tractable way to align Bayesian GSDs with accepted frequentist practice and regulatory expectations, supporting their robust application in confirmatory trials.
\end{abstract}

\begin{keyword}
Confirmatory trial \sep
Bayesian group sequential design \sep
Interim monitoring \sep
Early stopping \sep
Type I error control
\end{keyword}

\end{frontmatter}


\section{Introduction}
\label{sec:1}
Group sequential designs (GSDs) are the most widely used adaptive design in confirmatory clinical trials \citep{hatfield2016}. They enable interim analyses while preserving overall type I error control across interim and final analyses. By allowing researchers to evaluate accumulating trial data at prespecified points during the conduct of the trial, GSDs provide opportunities for early decision-making, such as stopping the trial for efficacy or futility when sufficient evidence has been gathered. GSDs are particularly valuable for improving trial efficiency by potentially saving time, costs and resources (\textit{i.e.}, around 68\% of published group sequential trials have been reported to stop early \citep{stevely2015}). Moreover, these designs improve ethical conduct by potentially reducing the number of participants exposed to ineffective or harmful treatments. 

Classical GSDs are a well-established approach in clinical trials \citep{jennison1999,wassmer2016} and are widely used in confirmatory studies \citep[\textit{e.g.},][]{keh2016,venkatesh2018}. These designs commonly involve null hypothesis testing at each interim analysis and calibrating the stopping boundaries over the interim analyses to preserve a nominal overall type I error rate \citep{haybittle1971,peto1976,pocock1977,obrien1979,lan1983,kim1987,hwang1990}. Their decision-making is based on whether predefined stopping boundaries are crossed. Such a framework offers flexibility and improves efficiency while ensuring robust statistical validity. In this context, these designs are referred to as frequentist GSDs.

\begin{figure}[!ht]
	\centering
	\includegraphics[width=\linewidth]{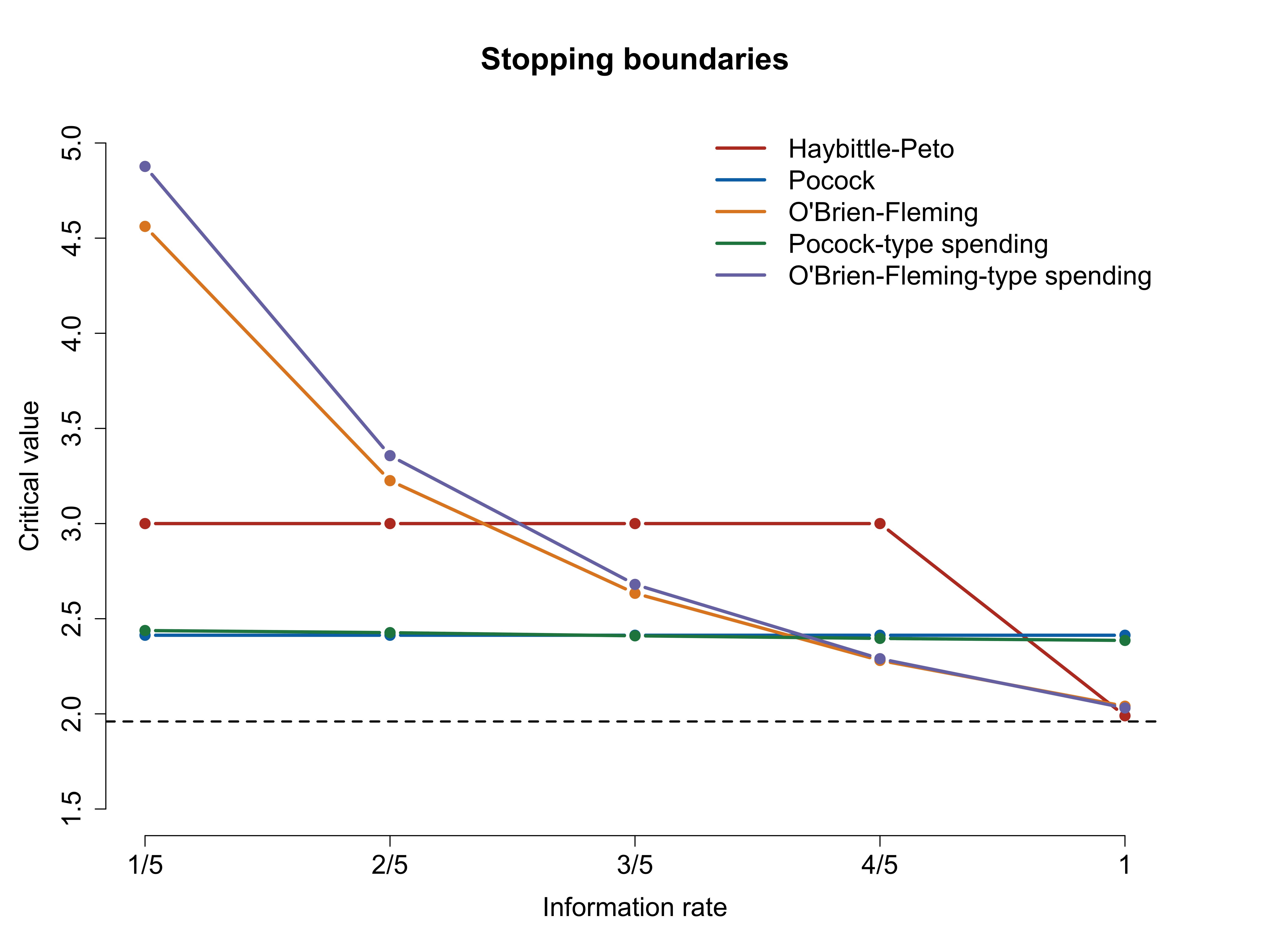}
	\caption{A comparison of common efficacy stopping boundaries in frequentist GSDs incorporating four equally spaced interim looks with a one-sided type I error rate of 0.025. 
	The Pocock- and O'Brien-Fleming-type spending is generated via the Lan-DeMets alpha-spending function. The information rate is the fraction of total information expected at the scheduled end of the trial \citep{demets1994}.}
	\label{fig:11}
\end{figure}

In frequentist GSDs, the most commonly used stopping boundaries are those introduced by O'Brien and Fleming \citep{obrien1979} and the Lan-DeMets error spending function that mimics O'Brien-Fleming-type properties \citep{lan1983}. These stopping rules are widely favoured, with about 61\% of published group sequential trials employing them \citep{stevely2015}. In contrast to alternatives such as the Pocock-type boundary \citep{pocock1977}, which applies a fixed threshold across all analyses, the O'Brien-Fleming-type boundary is deliberately conservative at early looks (\textit{i.e.}, requiring stronger evidence for early efficacy termination) while becoming progressively less stringent at later stages (see Figure~\ref{fig:11} for a comparison of commonly used stopping boundaries). This conservatism reduces the risk of overestimating the treatment effect and falsely concluding efficacy at early looks due to random fluctuations and noise, which aligns well with regulatory guidelines to ensure robust and reliable conclusions \citep{fda2019}.

Over the past two decades, there has been a growing exploration of GSDs within the Bayesian paradigm \citep[see][and references therein]{zhou2024,lee2024}, which we refer to as Bayesian GSDs. These designs provide a flexible and robust framework for leveraging prior information and incorporating accumulating evidence during the conduct of the trial. Unlike the frequentist method, Bayesian GSDs evaluate the probability of hypotheses or parameters of interest directly, allowing for a more intuitive and adaptive decision-making process. Specifically, prior knowledge about the treatment effect, often derived from historical information from completed trials and clinical opinions from published resources, is formalised into a prior distribution. As the trial progresses, this prior is updated with accumulating trial data at each interim analysis to generate a posterior distribution that reflects an up-to-date summary of the current understanding of the treatment effect. These posteriors form the basis for decisions on whether to stop the trial early for efficacy or futility \citep{spiegelhalter2004,berry2010}, utilising metrics such as decisions based on posterior probabilities, predictive probabilities or Bayes factors \citep[see][for detailed discussion]{gsponer2014,saville2014,lee2024}. Controlling the overall type I error rate is also crucial for Bayesian GSDs to align with regulatory standards, ensuring ethical trial conduct and preserving the credibility and reliability of trial conclusions \citep{shi2019}, although its necessity remains a topic of ongoing debate \citep{ryan2020b}.

In current practice, however, Bayesian GSDs are most often implemented using fixed posterior probability thresholds applied uniformly at all interim analyses \citep[\textit{e.g.},][]{reardon2017,jansen2023}. Specifically, efficacy or futility is declared when the posterior probability of the treatment effect exceeding (or failing to exceed) a prespecified effect threshold crosses a predefined probability cutoff (\textit{e.g.}, \citet{gsponer2014} recommend choosing an effect threshold of zero together with a probability cutoff equal to one minus the nominal type I error rate for efficacy monitoring). For example, efficacy may be declared whenever the posterior probability that the treatment effect exceeds a clinically meaningful threshold crosses a prespecified cutoff, such as 0.975 or 0.990, regardless of the timing of the analysis. Although these thresholds are usually calibrated to control the overall type I error rate, they do not depend on the amount of information accrued at the interim look.

However, such conventional Bayesian GSDs exhibit a subtle inconsistency compared to well-established frequentist procedures such as O'Brien-Fleming designs: whereas the latter explicitly penalise early analyses by requiring much stronger evidence at low information fractions, fixed posterior probability thresholds in Bayesian GSDs treat all interim analyses equally. As a result, Bayesian GSDs implemented in this conventional manner tend to allow much earlier stopping for efficacy, driven by greater sensitivity to random fluctuations in small samples. This behaviour is not a property of Bayesian inference itself, but rather of the specific decision rules that are most commonly used in practice.

Despite its importance for scientific validity, regulatory acceptability and the stability of treatment effect estimates, this discrepancy between Bayesian and frequentist early stopping pattern has received limited systematic attention. Aggressive early stopping is known to inflate observed treatment effects and undermine the reliability of confirmatory evidence, a concern that motivates the widespread regulatory preference for conservative early stopping boundaries such as O'Brien-Fleming \citep{pocock1999,fda2019}. However, conventional Bayesian GSDs, even when calibrated to control the same overall type I error, can recommend stopping at interim analyses where a frequentist O'Brien-Fleming boundary would still be far from being crossed.

In this work, we examine and address this inconsistency using the HYdrocortisone for PREvention of Septic Shock (HYPRESS) trial as a motivating case study \citep{keh2016}. We show that a conventional Bayesian GSD with fixed posterior probability thresholds behaves more like a Pocock design than an O'Brien-Fleming design, leading to substantially more aggressive early stopping. We therefore propose two practically implementable refinements to Bayesian interim monitoring that restore conservative early stopping while preserving the interpretability and flexibility of Bayesian decision-making. The first introduces a two-phase structure for posterior probability thresholds, and the second replaces posterior with predictive probability monitoring at interim looks. Both require only one additional tuning parameter and can be efficiently calibrated by the BaySeq simulation-free framework \citep{he2026a}. Collectively, these methods align Bayesian GSDs with the conservative early stopping behaviour underpinning frequentist practice and regulatory guidance.

\section{Aggressive Bayesian vs conservative frequentist GSDs}
\label{sec:2}
In this section, we first provide an overview of the HYPRESS trial and then use it as a case study to illustrate the contrasting early stopping behaviours of frequentist and Bayesian GSDs, highlighting their differences and practical implications.

\subsection{The HYPRESS trial}
\label{sec:21}
The HYPRESS trial was a placebo-controlled, double-blind, randomised clinical trial conducted in 34 centres in Germany to test the primary hypothesis that the early use of hydrocortisone compared to placebo reduces progression to septic shock in patients with severe sepsis. The primary outcome was the development of septic shock within 14 days. Given a 40\% 14-day occurrence of septic shock in the placebo group, HYPRESS aimed to detect a 15\% reduction in the hydrocortisone group, with a two-sided type I error rate of 5\% and 80\% power. HYPRESS adopted a frequentist three-stage O'Brien-Fleming GSD \citep{obrien1979} with equal allocation and two efficacy interim looks conducted after approximately one-third and two-thirds of the planned total enrolment had accrued primary outcome data. The stopping boundary was implemented through the Lan-DeMets alpha-spending function \citep{lan1983}. Further details of the trial design and conduct are provided in \citet{keh2016}.

\subsection{Contrasting early stopping tendencies}
\label{sec:22}
To demonstrate contrasting early stopping tendencies of Bayesian and frequentist GSDs, we redesign the HYPRESS trial under a Bayesian GSD incorporating a single posterior probability criterion for efficacy. The probability threshold is held constant across all analyses and calibrated to maintain the nominal overall type I error rate. We assume no prior knowledge of the treatment effect, therefore adopting a uniform prior for the event probability of each group in the Bayesian GSD. Furthermore, we specify a one-sided type I error rate of 2.5\%, rather than the two-sided 5\% used in the original HYPRESS trial, due to the common preference for one-sided superiority testing in drug development programs \citep{grieve2024}. All other design specifications are retained consistent with those of the original HYPRESS trial.

\begin{figure}[!ht]
	\centering
	\includegraphics[width=\linewidth]{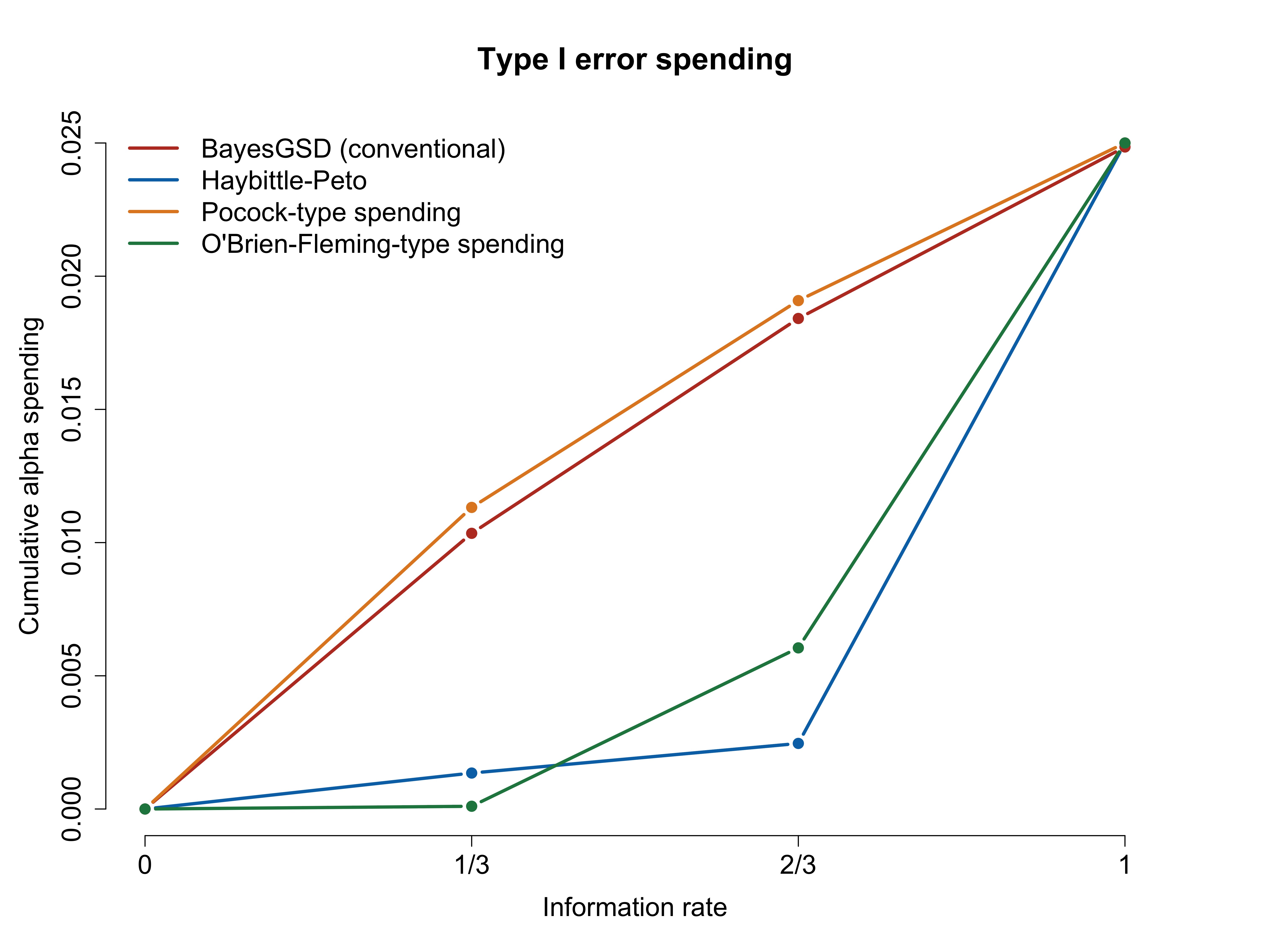}
	\caption{The cumulative alpha-spending profiles for the conventional Bayesian GSD and widely used frequentist GSDs incorporating two equally spaced interim analyses with a one-sided type I error rate of 2.5\%. 
	The Pocock- and O'Brien-Fleming-type spending is generated via the Lan-DeMets alpha-spending function.}
	\label{fig:23}
\end{figure}

We present the cumulative error spent at each analysis for the Bayesian GSD in Figure~\ref{fig:23}, where the efficacy stopping criterion is specified as the posterior probability that hydrocortisone has a lower 14-day occurrence of septic shock exceeding 0.9890, examined at each interim look and the final analysis, with a total enrolment target of 356 patients. The minimum total sample size of 356 patients and a probability threshold of 0.9890 are calibrated through \citet{he2026a} to maintain a one-sided type I error rate of 0.025 and 80\% power (see Figures~\ref{fig:21} and \ref{fig:22}). For comparison, we also illustrate the corresponding cumulative alpha spending for the Haybittle-Peto \citep{haybittle1971,peto1976}, Pocock \citep{pocock1977} and O'Brien-Fleming \citep{obrien1979} GSDs. The associated operating characteristics are summarised in Table~\ref{tab:21}. Unless otherwise stated, the Pocock and O'Brien-Fleming stopping boundaries are approximated through the Lan-DeMets alpha spending function \citep{lan1983}.

From Figure~\ref{fig:23}, the cumulative type I error spending of the Bayesian GSD incorporating a constant posterior probability threshold for efficacy across all analyses closely mirrors that of the Pocock GSD. Both designs exhibit nearly identical operating characteristics (see Table~\ref{tab:21}). In contrast to the Haybittle-Peto and O'Brien-Fleming GSDs, the Bayesian GSD shows a markedly higher probability of early efficacy stopping, requiring less extreme evidence to stop early. More specifically, under the null hypothesis, the probability of stopping for efficacy at the first interim analysis is 0.0103 for the Bayesian GSD, compared with 0.0013 for the Haybittle-Peto GSD and 0.0001 for the O'Brien-Fleming GSD. This pattern highlights the substantially more aggressive early stopping behaviour inherent to the Bayesian GSD, achieving greater early power (29.43\%) to detect the assumed treatment effect but at the cost of an order-of-magnitude to two-orders-of-magnitude increase in early type I error.

\section{Strategies to transition from aggressive to conservative early stopping}
\label{sec:3}
As shown in Section~\ref{sec:22}, Bayesian GSDs often exhibit more aggressive early stopping when equally spaced interim analyses are combined with the widely used posterior-probability decision rule employing fixed thresholds. This behaviour arises largely from the sensitivity of posterior probabilities to early data fluctuations and limited sample sizes. Such aggressive early stopping, however, is generally discouraged in group sequential trials because of scientific, regulatory and ethical concerns \citep{pocock1999}. To address this challenge, in this section we introduce two practical strategies that moderate overly aggressive early stopping in conventional Bayesian GSDs by either refining posterior-probability thresholds or incorporating predictive-probability criteria, therefore bringing Bayesian monitoring rules more in line with established frequentist GSD practice.
Based on the HYPRESS trial, assuming a septic shock rate of $\vartheta_{0}=0.40$ in the placebo group and $\vartheta_{1}=0.25$ in the hydrocortisone group, we consider the hypothesis test $H_{0}:\vartheta_{1}-\vartheta_{0} \geq 0$ versus $H_{1}:\vartheta_{1}-\vartheta_{0}<0$, with a one-sided type I error rate of $\alpha=0.025$ and 80\% power. Patients are randomised in equal proportions to the placebo and hydrocortisone groups.
In the Bayesian GSD, for interim looks, we define the posterior probability based stopping criterion for efficacy as
\begin{linenomath}
	\begin{equation}
		\label{eqn:3001}
		\Pr(\vartheta_{1}-\vartheta_{0}<0 \mid i_{k},j_{k};N_{k})>p_{k}
	\end{equation}
\end{linenomath}
and the predictive probability based stopping criterion for efficacy as 
\begin{linenomath}
	\begin{equation}
		\label{eqn:3002}
		\Pr(\Pr(\vartheta_{1}-\vartheta_{0}<0 \mid i_{K},j_{K};N_{K})>p_{K} \mid i_{k},j_{k};N_{k})>q_{k},
	\end{equation}
\end{linenomath}
respectively, for $k=1,2,\ldots,K-1$, where $i_{k}$ and $j_{k}$ represent the number of patients who develop septic shock in the placebo and hydrocortisone groups, respectively, at the $k$-th analysis, $p_{k}$ and $q_{k}$ are the posterior probability threshold and predictive probability threshold, respectively, at the $k$-th analysis, $N_{k}$ is the cumulative sample size at the $k$-th analysis, and $K$ denotes the total number of analyses. For the final analysis, the decision criterion for efficacy is 
\begin{linenomath}
	\begin{equation}
		\label{eqn:3003}
		\Pr(\vartheta_{1}-\vartheta_{0}<0 \mid i_{K},j_{K};N_{K})>p_{K}.
	\end{equation}
\end{linenomath}
To simplify implementation and facilitate interpretation, conventional Bayesian GSDs typically adopt a single fixed posterior probability based efficacy criterion across all equally spaced looks, enforcing $p_{k}=p$ for $k=1,2,\ldots,K$. Calibration of the decision criteria and evaluation of the operating characteristics are conducted using BaySeq, a simulation-free framework for efficient and exact assessment of Bayesian GSD operating characteristics \citep{he2026a}.


In evaluating the operating characteristics of the proposed strategies, we include four equally spaced interim analyses at 20\%, 40\%, 60\% and 80\% of the planned total enrolment, based on the availability of primary outcome data. We fix the maximum total sample size at 368 patients, which is the smallest total sample size to achieve the target type I error rate and power under a Bayesian GSD with a probability threshold of $p=0.9920$ applied uniformly across all analyses (see Figures~\ref{fig:301} and \ref{fig:302}). This conventional Bayesian design serves as the baseline comparator. For each strategy, the probability thresholds are calibrated to maintain control of the overall type I error rate. We then evaluate and compare the resulting power, stage-wise probabilities of stopping for efficacy and expected sample sizes under both the null and alternative hypotheses. 


\subsection{Strategy 1: replacing fixed with two-phase probability thresholds}
\label{sec:32}
The first strategy consists in partitioning the trial into early and late phases according to the information fraction (\textit{i.e.}, the proportion of the total information), such that $t<t^{*}$ corresponds to the early phase and $t \geq t^{*}$ to the late phase, where $t$ and $t^{*}$ are expressed as information fractions. Like the conventional Bayesian GSD, we adopt a single posterior probability based stopping criterion across all analyses but impose probability thresholds $p_{k}=p^{e}$ for $t_{k}<t^{*}$ and $p_{k}=p^{l}$ for $t_{k} \geq t^{*}$, where $p^{e}$ and $p^{l}$ represent the posterior probability thresholds for efficacy in the early and late phases, respectively. Specifically, the efficacy decision rule can be formulated as
\begin{linenomath}
	\begin{equation}
		\label{eqn:3201}
		\left\{
    \begin{aligned}
			\Pr(\vartheta_{1}-\vartheta_{0}<0 \mid i_{k},j_{k};N_{k})>p^{e}, & \text{ for $k \in \{k=1,2,\ldots,K \mid t_{k}<t_{*}\}$} \\
			\Pr(\vartheta_{1}-\vartheta_{0}<0 \mid i_{k},j_{k};N_{k})>p^{l}, & \text{ for $k \in \{k=1,2,\ldots,K \mid t_{k} \geq t_{*}\}$}.
    \end{aligned}
  	\right.
	\end{equation}
\end{linenomath}
To achieve conservative early stopping behaviour in the Bayesian GSD, we can select a large $t^{*}$ and a sufficiently stringent $p^{e}$ for the posterior probability based stopping criterion in the early phase and calibrate $p^{l}$ (necessarily with $p^{l}<p^{e}$) to preserve the desired overall type I error rate. 

\begin{figure}[!ht]
	\centering
	\includegraphics[width=\linewidth]{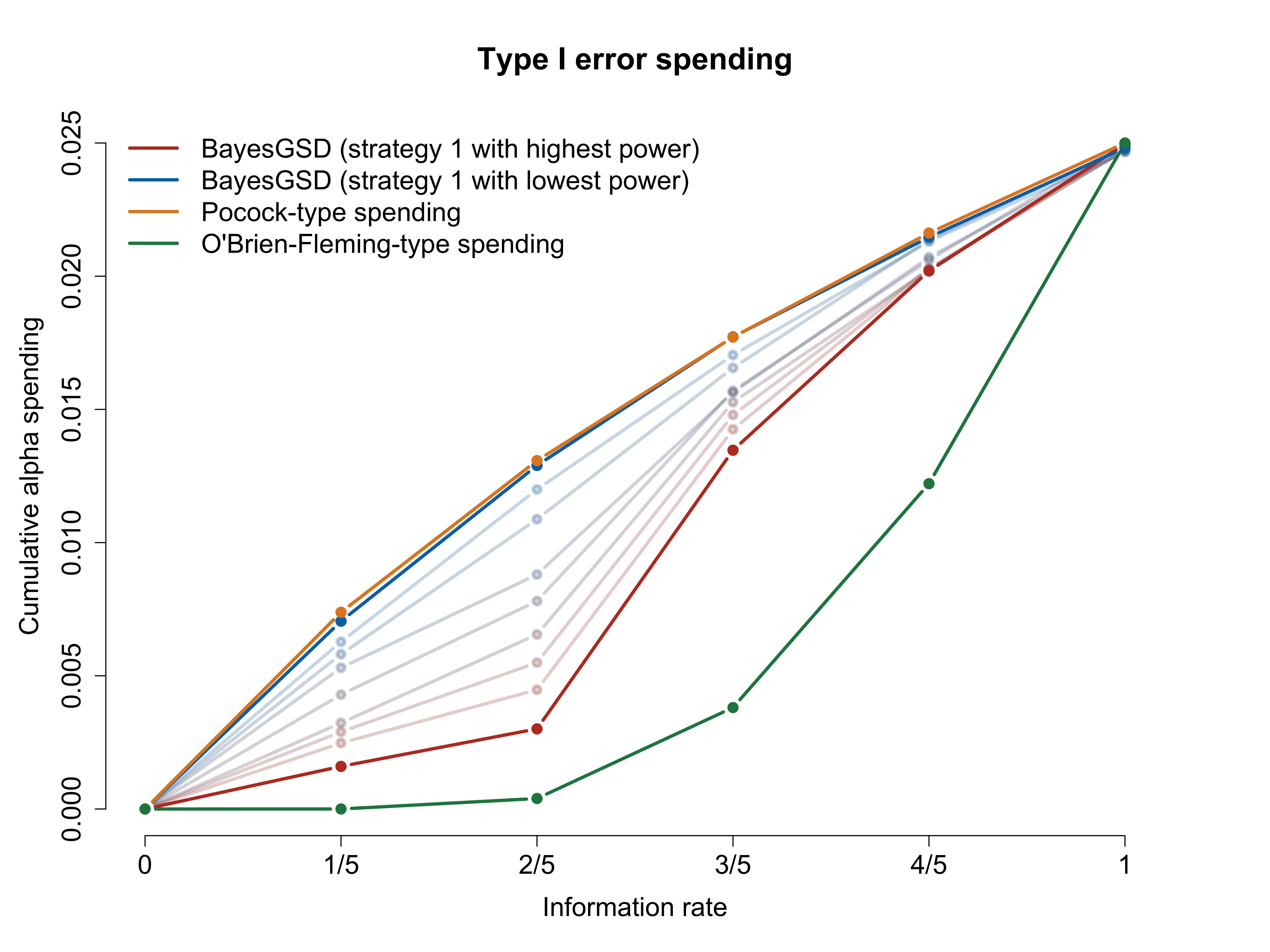}
	\caption{The cumulative alpha-spending profiles for the calibrated Bayesian GSDs under Strategy 1. The Colour gradient reflects the associated power of each calibrated design: warmer (red) shades correspond to higher power, whereas cooler (blue) shades indicate lower power. The Pocock- and O'Brien-Fleming-type spending is generated via the Lan-DeMets alpha-spending function.}
	\label{fig:325}
\end{figure}

To determine probability thresholds that yield the desired operating characteristics, we fix $t^{*}=0.50$ and allow $p^{e}$ and $p^{l}$ to vary over $[0.980,0.999]$ on a grid with increments of 0.0005. For each candidate combination $(p^{e},p^{l})$, we calculate the operating characteristics of the Bayesian GSD defined by the decision criteria in Eq.~(\ref{eqn:3201}). The resulting overall type I error rates, powers and expected sample sizes under the null and alternative hypotheses are shown in Figures~\ref{fig:321}--\ref{fig:324}, respectively. For each value of $p^{l}$, we then choose the smallest $p^{e}$ that maintains control of the overall type I error rate at 0.025 while maximising power.
Finally, we restrict attention to calibrated pairs that achieve at least 80\% power. The resulting calibrated $(p^{e},q^{l})$ pairs, together with their corresponding operating characteristics, are summarised in Table~\ref{tab:321}. Figure~\ref{fig:325} displays the corresponding cumulative alpha-spending profiles, presented alongside those for the Pocock and O'Brien-Fleming GSDs. The corresponding cumulative beta-spending profiles are provided in Figures~\ref{fig:326}.

\begin{table}[!ht]
	\centering
	\begin{tabular}{ccccccc}
		\toprule
		Threshold        & Type I error & Power   & \multicolumn{2}{c}{$\Pr(\text{early stop})$} & \multicolumn{2}{c}{Expected sample size} \\
		\cline{4-5} 
		\cline{6-7}
		$(p^{e},p^{l})$  &              &         & $H_{0}$ & $H_{1}$                            & $H_{0}$ & $H_{1}$                        \\
		\hline
		(0.9982, 0.9880) & 0.02495      & 0.83358 & 0.02020 & 0.72723                            & 365.208 & 256.349                        \\
		(0.9977, 0.9885) & 0.02488      & 0.82948 & 0.02025 & 0.72348                            & 364.973 & 254.331                        \\
		(0.9967, 0.9890) & 0.02474      & 0.82704 & 0.02029 & 0.72063                            & 364.824 & 251.358                        \\
		(0.9962, 0.9895) & 0.02475      & 0.82204 & 0.02024 & 0.71272                            & 364.688 & 250.911                        \\
		(0.9957, 0.9900) & 0.02499      & 0.81853 & 0.02060 & 0.70844                            & 364.460 & 249.359                        \\
		(0.9952, 0.9905) & 0.02469      & 0.81447 & 0.02071 & 0.70522                            & 364.308 & 248.205                        \\
		(0.9938, 0.9910) & 0.02486      & 0.80978 & 0.02137 & 0.70040                            & 364.001 & 245.877                        \\
		(0.9927, 0.9915) & 0.02466      & 0.80552 & 0.02130 & 0.69559                            & 363.853 & 243.027                        \\
		(0.9920, 0.9920) & 0.02482      & 0.80093 & 0.02146 & 0.69250                            & 363.667 & 241.840                        \\
		\bottomrule
	\end{tabular}%
	\caption{The resulting calibrated probability thresholds and their corresponding operating characteristics under Strategy 1.}
	\label{tab:321}
\end{table}

Table~\ref{tab:321} illustrates that, for a fixed sample size, tightening the early-phase threshold while relaxing the late-phase threshold leads to a monotonic increase in power, ranging from 80.09\% for the conventional Bayesian GSD (0.9920,0.9920) to 83.36\% for the most asymmetric threshold pair (0.9982,0.9880). Across all calibrated designs, the overall type I error rate remains tightly controlled around the nominal 0.025 level (0.02466--0.02499). The probability of early stopping under the null remains consistently low at about 2\%, whereas the probability of early stopping under the alternative increases from 69.3\% to 72.7\% as the early-phase threshold becomes more stringent and the late-phase threshold becomes more permissive. Expected sample sizes exhibit a modest upward trend under the alternative (from 242 to 256 subjects) while under the null they remain close to the maximum of 368. Importantly, calibrated designs achieve higher power than the conventional Bayesian GSD but at the cost of requiring a larger expected sample size under the alternative, reflecting the more conservative early stopping behaviour induced by Strategy 1.

The apparent discrepancy between the increase in early stopping probability and the modest rise in expected sample size under the alternative arises from how early- and late-phase decisions interact. More stringent early-phase thresholds reduce early stopping at the first two looks but increase stopping at later interim analyses (see Figure~\ref{fig:326}). Although this increases the overall early stopping probability, shifting decisions from early to late interim analyses raises the number of patients accrued before stopping, therefore yielding a slightly larger expected sample size. 

Figure~\ref{fig:325} illustrates the cumulative alpha-spending patterns for the calibrated designs. The strategies span a continuum between the conservative O'Brien-Fleming profile and the aggressive Pocock profile. In the early phase, the calibrated designs spend alpha at levels that fall between these two extremes, reflecting the desired moderation of early stopping behaviour. In the later phase, however, the spending profiles shift closer to the Pocock pattern, consistent with the use of more permissive late-phase thresholds. Notably, when the early-phase threshold is tightened, a sharper increase in alpha spending emerges at the transition point between the early and late phases, highlighting the intended shift from conservative to more liberal decision criteria once a sufficient amount of information has accrued. This illustrates that Strategy 1 can meaningfully reshape the early stopping behaviour of Bayesian GSDs: by appropriately calibrating early- and late-phase thresholds, the design produces alpha-spending profiles that more closely align with established frequentist principles while improving control of early false positives compared with conventional Bayesian GSDs.

\subsection{Strategy 2: replacing posterior with predictive probability criteria}
\label{sec:34}
The second strategy consists in replacing posterior probability stopping rules with predictive probability based criteria for interim monitoring in the conventional Bayesian GSD. More specifically, predictive probability is used to govern interim stopping, while posterior probability is retained for the final analysis. Under this hybrid framework, the efficacy decision rule can be written as
\begin{linenomath}
	\begin{equation}
		\label{eqn:3401}
		\left\{
    \begin{aligned}
			\Pr(\vartheta_{1}-\vartheta_{0}<0 \mid i_{k},j_{k};N_{k})>p, & \text{ for $k=K$} \\
			\Pr(\Pr(\vartheta_{1}-\vartheta_{0}<0 \mid i_{K},j_{K};N_{K})>p \mid i_{k},j_{k};N_{k})>q, & \text{ for $k<K$}.
    \end{aligned}
  	\right.
	\end{equation}
\end{linenomath}
We choose a sufficiently stringent $q$ for the predictive probability based stopping criterion at interim analyses and then calibrate $p$ to preserve the required overall type I error rate. Using a predictive probability based stopping criterion at interim looks naturally induces more conservative early stopping behaviour as the predictive probability accounts for uncertainty about unobserved future outcomes and therefore reduces sensitivity to random fluctuations in the early stage.

To determine probability thresholds that achieve the desired operating characteristics, we vary $p \in [0.975,0.992]$ and $q \in [0.825,0.992]$ over a predefined grid (using increments of 0.0005 for both) and evaluate the Bayesian GSD defined with the decision criteria in Eq.~(\ref{eqn:3401}) for every candidate pair $(p,q)$. The resulting overall type I error rates, powers, and expected sample sizes under the null and alternative hypotheses are shown in Figures~\ref{fig:341}--\ref{fig:344}. For each fixed value of $p$, we identify the smallest $q$ that maintains control of the overall type I error rate at 0.025 while maximising power. 
The resulting calibrated $(p,q)$ pairs and their corresponding operating characteristics are summarised in Table~\ref{tab:341}. Figure~\ref{fig:343} illustrates the cumulative alpha-spending profile for the selected designs, alongside those of the Pocock and O'Brien-Fleming GSDs, with the corresponding cumulative beta-spending profiles provided in Figure~\ref{fig:346}.

\begin{table}[!ht]
	\centering
	\begin{tabular}{ccccccc}
		\toprule
		Threshold        & Type I error & Power   & \multicolumn{2}{c}{$\Pr(\text{early stop})$} & \multicolumn{2}{c}{Expected sample size} \\
		\cline{4-5} 
		\cline{6-7}
		$(p,q)$          &              &         & $H_{0}$ & $H_{1}$                            & $H_{0}$ & $H_{1}$                        \\
		\hline
		(0.9760, 0.9920) & 0.02499      & 0.87322 & 0.00298 & 0.49128                            & 367.707 & 304.069                        \\
		(0.9765, 0.9864) & 0.02499      & 0.87104 & 0.00350 & 0.52484                            & 367.630 & 296.228                        \\
		(0.9770, 0.9858) & 0.02499      & 0.86917 & 0.00368 & 0.52689                            & 367.613 & 295.591                        \\
		(0.9775, 0.9851) & 0.02498      & 0.86719 & 0.00405 & 0.52803                            & 367.571 & 294.924                        \\
		(0.9780, 0.9808) & 0.02499      & 0.86561 & 0.00494 & 0.53787                            & 367.401 & 291.329                        \\
		(0.9785, 0.9735) & 0.02499      & 0.86485 & 0.00644 & 0.57641                            & 367.111 & 281.147                        \\
		(0.9790, 0.9623) & 0.02499      & 0.86476 & 0.00846 & 0.59950                            & 366.714 & 269.733                        \\
		(0.9795, 0.9531) & 0.02499      & 0.86446 & 0.01001 & 0.62687                            & 366.465 & 261.837                        \\
		(0.9800, 0.9458) & 0.02487      & 0.86380 & 0.01081 & 0.63622                            & 366.293 & 257.112                        \\
		(0.9805, 0.9433) & 0.02498      & 0.86243 & 0.01126 & 0.63749                            & 366.184 & 255.987                        \\
		(0.9810, 0.9416) & 0.02499      & 0.86051 & 0.01176 & 0.63448                            & 366.069 & 255.792                        \\
		(0.9815, 0.9402) & 0.02482      & 0.85820 & 0.01179 & 0.63311                            & 366.057 & 255.561                        \\
		(0.9820, 0.9389) & 0.02499      & 0.85645 & 0.01205 & 0.63323                            & 365.992 & 255.335                        \\
		(0.9825, 0.9359) & 0.02499      & 0.85347 & 0.01213 & 0.63298                            & 365.958 & 253.738                        \\
		(0.9830, 0.9266) & 0.02498      & 0.85346 & 0.01283 & 0.65392                            & 365.863 & 247.529                        \\
		(0.9835, 0.9183) & 0.02498      & 0.85297 & 0.01375 & 0.66098                            & 365.701 & 242.648                        \\
		(0.9840, 0.9090) & 0.02472      & 0.84954 & 0.01442 & 0.66372                            & 365.526 & 241.048                        \\
		(0.9845, 0.9030) & 0.02461      & 0.84867 & 0.01498 & 0.66728                            & 365.400 & 238.421                        \\
		(0.9850, 0.8956) & 0.02484      & 0.84755 & 0.01592 & 0.67313                            & 365.163 & 236.573                        \\
		(0.9855, 0.8938) & 0.02495      & 0.84547 & 0.01630 & 0.67319                            & 365.070 & 236.477                        \\
		(0.9860, 0.8938) & 0.02478      & 0.84238 & 0.01619 & 0.66860                            & 365.078 & 237.116                        \\		
		(0.9865, 0.8919) & 0.02498      & 0.83982 & 0.01667 & 0.66898                            & 365.032 & 236.937                        \\
		(0.9870, 0.8834) & 0.02484      & 0.83805 & 0.01707 & 0.67661                            & 364.986 & 233.268                        \\
		(0.9875, 0.8676) & 0.02499      & 0.83688 & 0.01796 & 0.68530                            & 364.760 & 227.769                        \\
		(0.9880, 0.8558) & 0.02498      & 0.83683 & 0.01862 & 0.69080                            & 364.602 & 224.042                        \\
		(0.9885, 0.8514) & 0.02462      & 0.83367 & 0.01857 & 0.69236                            & 364.598 & 223.290                        \\
		(0.9890, 0.8451) & 0.02497      & 0.83277 & 0.01953 & 0.69722                            & 364.459 & 221.220                        \\
		(0.9895, 0.8401) & 0.02499      & 0.82969 & 0.01976 & 0.69785                            & 364.440 & 220.851                        \\
		(0.9900, 0.8363) & 0.02455      & 0.82682 & 0.01935 & 0.69761                            & 364.480 & 220.779                        \\
		(0.9905, 0.8270) & 0.02442      & 0.82378 & 0.01966 & 0.69720                            & 364.356 & 219.867                        \\
		(0.9910, 0.8224) & 0.02493      & 0.81906 & 0.02072 & 0.69376                            & 363.965 & 220.133                        \\
		(0.9915, 0.8213) & 0.02444      & 0.81440 & 0.02048 & 0.68989                            & 363.986 & 220.608                        \\
		(0.9920, 0.8185) & 0.02439      & 0.80897 & 0.02054 & 0.68447                            & 363.886 & 221.565                        \\
		\bottomrule
	\end{tabular}%
	\caption{The resulting calibrated probability thresholds and their corresponding operating characteristics under Strategy 2.}
	\label{tab:341}
\end{table}

\begin{figure}[!ht]
	\centering
	\includegraphics[width=\linewidth]{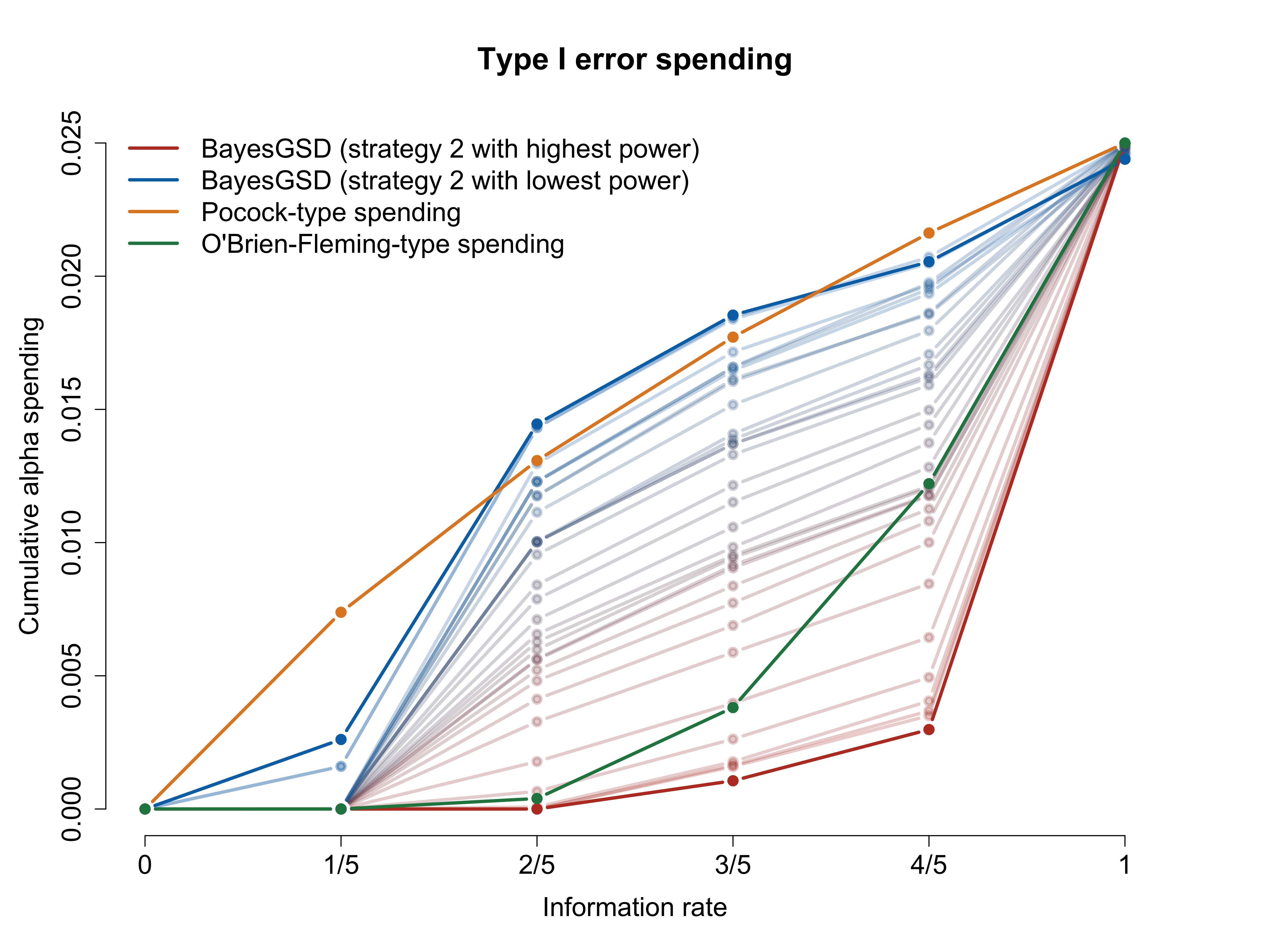}
	\caption{The cumulative alpha-spending profiles for the calibrated Bayesian GSDs under Strategy 2. The Colour gradient reflects the associated power of each calibrated design: warmer (red) shades correspond to higher power, whereas cooler (blue) shades indicate lower power. The Pocock- and O'Brien-Fleming-type spending is generated via the Lan-DeMets alpha-spending function.}
	\label{fig:345}
\end{figure}

As illustrated in Table~\ref{tab:341}, the overall type I error rate remains tightly controlled around the nominal 0.025 level (0.02439--0.02499) across all calibrated designs. As the posterior probability threshold $p$ increases and the predictive probability threshold 
$q$ becomes more permissive, power falls from 87.3\% for (0.9760,0.9920) to 80.90\% for (0.9920,0.8185). Early stopping remains low under the null, ranging from 0.30\% to 2.05\%, while under the alternative it varies from 49.13\% to 68.45\%. This wide range of early stopping probabilities highlights the flexibility of Strategy 2 in shaping interim decision behaviour, offering a broader spectrum of stopping patterns compared to Strategy 1. Expected sample sizes under the null remain close to the maximum of 368, while expected sample sizes under the alternative drop from about 304 to 222 as the design becomes more aggressive in the late phase. Notably, several calibrated designs achieve higher power than the conventional Bayesian GSD while simultaneously requiring a smaller expected sample size under the alternative, demonstrating improved efficiency relative to the baseline design.

Figure~\ref{fig:345} shows how Strategy 2 reshapes cumulative alpha spending. All calibrated designs closely track the O'Brien-Fleming profile at the earliest interim analysis, reflecting strong early conservatism induced by the predictive probability stopping rule. As information accumulates, the alpha-spending curves diverge from the O'Brien-Fleming trajectory across the intermediate interims depending on the chosen probability thresholds. The most conservative design spends even less alpha than O'Brien-Fleming, whereas the most aggressive design surpasses the Pocock boundary. Most calibrated designs, however, fall between these established profiles, illustrating that Strategy 2 enables a broad and flexible range of spending behaviours at intermediate looks while still maintaining tight control over early stopping.


\section{Discussion}
\label{sec:4}
In this work, we have examined an important but underappreciated inconsistency between conventional Bayesian and widely used frequentist GSDs. Although Bayesian GSDs offer substantial unique advantages, in particular their coherent probabilistic interpretation and ability to incorporate prior information \citep{spiegelhalter2004,berry2010}, our results show that the posterior probability stopping criteria most commonly used in practice, when applied with fixed thresholds across all interim analyses, can lead to substantially more aggressive early stopping than is typical in frequentist GSDs such as O'Brien-Fleming.

Importantly, the discrepancy does not arise from Bayesian principles themselves but from the specific form of the decision criteria that are typically used in practice. Stopping boundaries commonly employed in frequentist GSDs are explicitly constructed to be conservative at early information fractions, reflecting long-standing methodological principles \citep{jennison1999} and regulatory expectations \citep{fda2019} for confirmatory trials. In contrast, fixed probability thresholds in conventional Bayesian GSDs implicitly treat all interim looks as equally informative, even though early analyses are inherently unstable. As illustrated in the HYPRESS case study, this difference can translate into materially divergent decisions, even when evaluating the same accumulating evidence. 

Aggressive early stopping has important practical consequences beyond its theoretical implications. Premature stopping is well known to inflate treatment effect estimates and undermine the robustness of confirmatory evidence, which underpins the long-standing regulatory preference for conservative early stopping boundaries \citep{pocock1999,fda2019}. Consequently, Bayesian GSDs that exhibit Pocock-type behaviour rather than O'Brien-Fleming-type behaviour may warrant regulatory caution, even when calibrated to control the overall type I error rate.

To mitigate this discrepancy, we have proposed two pragmatic refinements to decision criteria in conventional Bayesian GSDs that moderate overly aggressive early stopping. The first method divides the trial into early and late phases, applying more stringent probability thresholds during the early phase to delay early stopping and less stringent thresholds during the later phase to ensure sufficient power. The transition point between early and late phases can be chosen based on information accrual, with values around 40--60\% providing a natural shift from conservative to more permissive stopping. The second approach replaces posterior probability based stopping rules with predictive probability based decision criteria at interim looks to reduce sensitivity to random fluctuations during the early stage. Compared with conventional Bayesian GSDs, only a single additional tuning parameter is introduced in both procedures, and the designs remain straightforward to implement and calibrate, keeping computation practical. 

As the HYPRESS case study shows, both refinements effectively mitigate overly aggressive early stopping in Bayesian GSDs, although they differ in their respective trade-offs in statistical efficiency and flexibility in shaping early stopping pattern. The two-phase posterior probability refinement enables direct control over when the design shifts from conservative to more permissive evidence requirements, therefore allowing explicit specification of the information fraction at which additional alpha should begin to be released. The predictive probability refinement, by contrast, defers early stopping in a more organic manner: as it incorporates uncertainty about future unobserved outcomes, it naturally withholds alpha spending till a meaningful amount of information has accumulated (\textit{i.e.}, in the case study, not until 40\% of the data were available) without requiring a predefined transition point. Although predictive probabilities do not permit fine-tuning through a single fixed threshold, they offer greater flexibility in shaping the trade-off between power, expected sample size and the overall stopping profile, enabling more adaptable control of design behaviour across a broad range of operating characteristics. Furthermore, the predictive probability refinement enables completely concave alpha-spending curves, analogous to O'Brien-Fleming boundaries, consistent with the principle that increasing information should permit proportionally greater alpha spending. This suggests that predictive probability monitoring offers a particularly appealing Bayesian analogue of frequentist information-dependent stopping boundaries.

Several authors have noted the close connection between Bayesian and frequentist GSDs, and in particular that conservative frequentist stopping boundaries can be reproduced in a Bayesian framework through the use of sceptical or robust priors \citep{stallard2020}. This suggests that, in principle, the degree of conservatism in Bayesian interim monitoring could be controlled via the choice of prior. In confirmatory trials, however, priors are typically intended to represent genuine external information, such as historical evidence, mechanistic understanding or expert opinion \citep{spiegelhalter2004}, and are therefore usually specified on scientific rather than operational grounds. Adjusting the prior primarily to achieve desired operating characteristics can conflate the distinction between evidence synthesis and design calibration and complicate the interpretation of the resulting posterior, potentially biasing posterior estimates and raising regulatory concerns. We therefore focus on modifying the interim decision criteria themselves, rather than tuning the prior specification, as a more transparent and principled way to achieve conservative early stopping behaviour in Bayesian GSDs.

In this work, we restrict attention to settings involving non-informative or weakly informative priors, representing common practice in confirmatory trials where prior evidence is limited or where neutrality is preferred. When genuinely informative priors are available and scientifically justified, their interaction with interim stopping rules may meaningfully influence early stopping behaviour, potentially amplifying or attenuating aggressiveness depending on the strength of the prior and its alignment with the true treatment effect. A systematic study of these interactions remains an important avenue for future research.

In most published Bayesian GSDs, interim monitoring relies on fixed posterior probability thresholds, largely for computational feasibility. Allowing probability thresholds to vary across interim looks or incorporating predictive probability based decision criteria while preserving type I and type II error constraints would dramatically increase the number of design parameters requiring calibration. In designs with multiple interim analyses and stopping rules, this rapidly results in prohibitive computational demands. Recent advances in computational methodology and software \citep{zhu2017a,zhu2017b,shi2019,couturier2024a,he2026a,he2026b} have begun to circumvent these challenges. In particular, \citet{he2026a} introduced BaySeq, a simulation-free framework that enables exact evaluation of Bayesian GSDs, making it feasible in practice to explore and implement more flexible and sophisticated decision criteria, including hybrid rules based on posterior and predictive probabilities. This overcomes a major historical obstacle to the practical adoption of information-dependent Bayesian monitoring rules in confirmatory trials.
In conclusion, conventional Bayesian GSDs based on fixed posterior probability thresholds can show overly aggressive early stopping that is misaligned with established frequentist practice and regulatory expectations. The two refinements proposed here offer tractable, principled and regulator-compatible ways to restore conservative early monitoring in Bayesian GSDs. By aligning Bayesian interim decision-making more closely with information-based stopping principles, these approaches strengthen the credibility, robustness and practical applicability of Bayesian designs in confirmatory trials. In practice, we further recommend that any proposed Bayesian GSD be accompanied by an evaluation of the implied cumulative alpha spending and the probability of stopping at each interim analysis, with these quantities visualised and compared against well-understood frequentist benchmarks prior to trial commitment, to ensure transparent and regulator-aligned control of early stopping behaviour.



\subsection*{Data availability statement}
The authors declare that all data necessary to confirm the conclusions of the present work are fully included within the article. The source code for evaluating the operating characteristics of the Bayesian GSDs discussed herein is publicly available at \url{https://github.com/zhangyi-he/BayesGSD}.

\subsection*{Funding statement}
The authors declare no funding associated with the work presented in this article. 

\subsection*{Conflict of interest disclosure}
The authors declare no potential conflict of interests.

\bibliography{ZH2024_Manuscript}

@article{hatfield2016,
  author={Hatfield, I and Allison, A and Flight, L and Julious, S A and Dimairo, M},
  title={{Adaptive designs undertaken in clinical research: a review of registered clinical trials}},
  journal={Trials},
  volume={17},
  number={1},
  pages={1--13},
  year={2016}
}

@article{stevely2015,
  author={Stevely, A and Dimairo, M and Todd, S and Julious, S A and Nicholl, J and others},
  title={{An investigation of the shortcomings of the CONSORT 2010 statement for the reporting of group sequential randomised controlled trials: a methodological systematic review}},
  journal={PLoS One},
  volume={10},
  number={11},
  pages={e0141104},
  year={2015}
}

@book{jennison1999,
  author={Jennison, C and Turnbull, B W},
  title={{Group Sequential Methods with Applications to Clinical Trials}},
  publisher={Chapman \& Hall/CRC},
  address={Boca Raton},
  year={1999}
}

@book{wassmer2016,
  author={Wassmer, G and Brannath, W},
  title={{Group Sequential and Confirmatory Adaptive Designs in Clinical Trials}},
  publisher={Springer},
  address={Heidelberg},
  year={2016}
}

@article{keh2016,
  author={Keh, D and Trips, E and Marx, G and Wirtz, S P and Abduljawwad, E and others},
  title={{Effect of hydrocortisone on development of shock among patients with severe sepsis: the HYPRESS randomized clinical trial}},
  journal={JAMA},
  volume={316},
  number={17},
  pages={1775--1785},
  year={2016}
}

@article{venkatesh2018,
  author={Venkatesh, B and Finfer, S and Cohen, J and Rajbhandari, D and Arabi, Y and others},
  title={{Adjunctive glucocorticoid therapy in patients with septic shock}},
  journal={New England Journal of Medicine},
  volume={378},
  number={9},
  pages={797--808},
  year={2018}
}

@article{haybittle1971,
  author={Haybittle, J L},
  title={{Repeated assessment of results in clinical trials of cancer treatment}},
  journal={The British Journal of Radiology},
  volume={44},
  number={526},
  pages={793--797},
  year={1971}
}

@article{peto1976,
  author={Peto, R and Pike, M C and Armitage, P and Breslow, N E and Cox, D R and others},
  title={{Design and analysis of randomized clinical trials requiring prolonged observation of each patient. I. Introduction and design}},
  journal={British Journal of Cancer},
  volume={34},
  number={6},
  pages={585--612},
  year={1976}
}

@article{pocock1977,
  author={Pocock, S J},
  title={{Group sequential methods in the design and analysis of clinical trials}},
  journal={Biometrika},
  volume={64},
  number={2},
  pages={191--199},
  year={1977}
}

@article{obrien1979,
  author={O'Brien, P C and Fleming, T R},
  title={{A multiple testing procedure for clinical trials}},
  journal={Biometrics},
  volume={35},
  number={3},
  pages={549--556},
  year={1979}
}

@article{lan1983,
  author={Lan, K K G and DeMets, D L},
  title={{Discrete sequential boundaries for clinical trials}},
  journal={Biometrika},
  volume={70},
  number={3},
  pages={659--663},
  year={1983}
}

@article{kim1987,
  author={Kim, K and DeMets, D L},
  title={{Design and analysis of group sequential tests based on the type I error spending rate function}},
  journal={Biometrika},
  volume={74},
  number={1},
  pages={149--154},
  year={1987}
}

@article{hwang1990,
  author={Hwang, I K and Shih, W J and De Cani, John S},
  title={{Group sequential designs using a family of type I error probability spending functions}},
  journal={Statistics in Medicine},
  volume={9},
  number={12},
  pages={1439--1445},
  year={1990}
}

@misc{fda2019,
  author={{US Food and Drug Administration}},
  title={{Adaptive Design Clinical Trials for Drugs and Biologics}},
  year={2019},
  note={Last accessed 16 June 2024},
  url={https://www.fda.gov/media/78495/download}
}

@article{demets1994,
  author={DeMets, D L and Lan, KK G},
  title={{Interim analysis: the alpha spending function approach}},
  journal={Statistics in Medicine},
  volume={13},
  number={13-14},
  pages={1341--1352},
  year={1994}
}

@article{zhou2024,
  author={Zhou, T and Ji, Y},
  title={{On Bayesian sequential clinical trial designs}},
  journal={The New England Journal of Statistics in Data Science},
  volume={2},
  number={1},
  pages={136--151},
  year={2024}
}

@article{lee2024,
  author={Lee, S Y},
  title={{Using Bayesian statistics in confirmatory clinical trials in the regulatory setting: a tutorial review}},
  journal={BMC Medical Research Methodology},
  volume={24},
  number={1},
  pages={110},
  year={2024}
}

@book{spiegelhalter2004,
  author={Spiegelhalter, D J and Abrams, K R and Myles, J P},
  title={Bayesian Approaches to Clinical Trials and Health-care Evaluation},
  publisher={Wiley},
  address={New York},
  year={2004}
}

@book{berry2010,
  author={Berry, S M and Carlin, B P and Lee, J J and Muller, P},
  title={Bayesian Adaptive Methods for Clinical Trials},
  publisher={Chapman \& Hall/CRC},
  address={London},
  year={2010}
}

@article{gsponer2014,
  author={Gsponer, T and Gerber, F and Bornkamp, B and Ohlssen, D and Vandemeulebroecke, M and others},
  title={{A practical guide to Bayesian group sequential designs}},
  journal={Pharmaceutical Statistics},
  volume={13},
  number={1},
  pages={71--80},
  year={2014}
}

@article{saville2014,
  author={Saville, B R and Connor, J T and Ayers, G D and Alvarez, J},
  title={{The utility of Bayesian predictive probabilities for interim monitoring of clinical trials}},
  journal={Clinical Trials},
  volume={11},
  number={4},
  pages={485--493},
  year={2014}
}

@article{shi2019,
  author={Shi, H and Yin, G},
  title={{Control of type I error rates in Bayesian sequential designs}},
  journal={Bayesian Analysis},
  volume={14},
  number={2},
  pages={399--425},
  year={2019}
}

@article{ryan2020b,
  author={Ryan, E G and Brock, K and Gates, S and Slade, D},
  title={{Do we need to adjust for interim analyses in a Bayesian adaptive trial design?}},
  journal={BMC Medical Research Methodology},
  volume={20},
  number={150},
  pages={1--9},
  year={2020}
}

@article{reardon2017,
  author={Reardon, M J and Van Mieghem, N M and Popma, J J and Kleiman, N S and S{\o}ndergaard, L and others},
  title={{Surgical or transcatheter aortic-valve replacement in intermediate-risk patients}},
  journal={New England Journal of Medicine},
  volume={376},
  number={14},
  pages={1321--1331},
  year={2017}
}

@article{jansen2023,
  author={Jansen, J O and Hudson, J and Cochran, C and MacLennan, G and Lendrum, R and others},
  title={{Emergency department resuscitative endovascular balloon occlusion of the aorta in trauma patients with exsanguinating hemorrhage: the UK-REBOA randomized clinical trial}},
  journal={JAMA},
  volume={330},
  number={19},
  pages={1862--1871},
  year={2023}
}

@article{pocock1999,
  author={Pocock, S and White, I},
  title={{Trials stopped early: too good to be true?}},
  journal={The Lancet},
  volume={353},
  number={9157},
  pages={943--944},
  year={1999}
}

@misc{he2026a,
  author={He, Z and Yu, F and Cro, S and Billot, L},
  title={{BaySeq: a simulation-free framework for exact evaluation of Bayesian group sequential designs}},
  year={2026},
  note={Manuscript submitted for publication}
}

@article{grieve2024,
  author={Grieve, A P},
  title={{Probability of success and group sequential designs}},
  journal={Pharmaceutical Statistics},
  volume={23},
  number={2},
  pages={185--203},
  year={2024}
}

@article{zhu2017a,
  author={Zhu, H and Yu, Q},
  title={{A Bayesian sequential design using alpha spending function to control type I error}},
  journal={Statistical Methods in Medical Research},
  volume={26},
  number={5},
  pages={2184--2196},
  year={2017}
}

@article{zhu2017b,
  author={Zhu, H and Yu, Q and Mercante, D E},
  title={{A Bayesian sequential design with binary outcome}},
  journal={Pharmaceutical Statistics},
  volume={16},
  number={3},
  pages={192--200},
  year={2017}
}

@article{couturier2024a,
  author={Couturier, D-L and Ryan, E G and Puhr, R and Jaki, T and Heritier, S},
  title={{A fast, flexible simulation framework for Bayesian adaptive designs--the R package BATSS}},
  journal={arXiv},
  pages={2410.02050},
  year={2024}
}

@misc{he2026b,
  author={He, Z and Yu, F and Cro, S and Billot, L},
  title={{Rapid evaluation of statistical and operational properties of Bayesian group sequential designs}},
  year={2026},
  note={Manuscript submitted for publication}
}

@article{stallard2020,
  author={Stallard, N and Todd, S and Ryan, E G and Gates, S},
  title={{Comparison of Bayesian and frequentist group-sequential clinical trial designs}},
  journal={BMC Medical Research Methodology},
  volume={20},
  number={4},
  pages={1--14},
  year={2020}
}

\clearpage
\appendix
\setcounter{figure}{0}
\setcounter{table}{0}

\section{Supplementary figures and tables}
\label{appx:1}

\vfill

\begin{table}[!ht]
	\centering
	\begin{tabular}{cccccc}
		\toprule
		Design        & Sample size & Type I error & Power   & \multicolumn{2}{c}{Expected sample size} \\
		\cline{5-6}
		              &             &              &         & $H_{0}$ & $H_{1}$                        \\
		\hline
		Bayesian GSD  & 356         & 0.02485      & 0.80628 & 352.6 & 250.0                            \\
		GSD with HP   & 306         & 0.02500      & 0.80000 & 305.6 & 270.9                            \\
		GSD with asP  & 356         & 0.02500      & 0.80000 & 351.9 & 248.9                            \\
		GSD with asOP & 308         & 0.02500      & 0.80000 & 307.0 & 262.9                            \\
		\bottomrule
	\end{tabular}%
	\caption{Operating characteristics for the conventional Bayesian GSD and for the GSDs incorporating the HP, asP and asOF stopping boundaries in the HYPRESS trial. HP represents the Haybittle-Peto GSD, and asP and asOF denote the Pocock and O'Brien-Fleming GSDs, respectively, where the stopping rules are implemented via the Lan-DeMets alpha-spending function.}
	\label{tab:21}
\end{table}

\vfill

\begin{figure}[p]
	\centering
	\includegraphics[width=\linewidth]{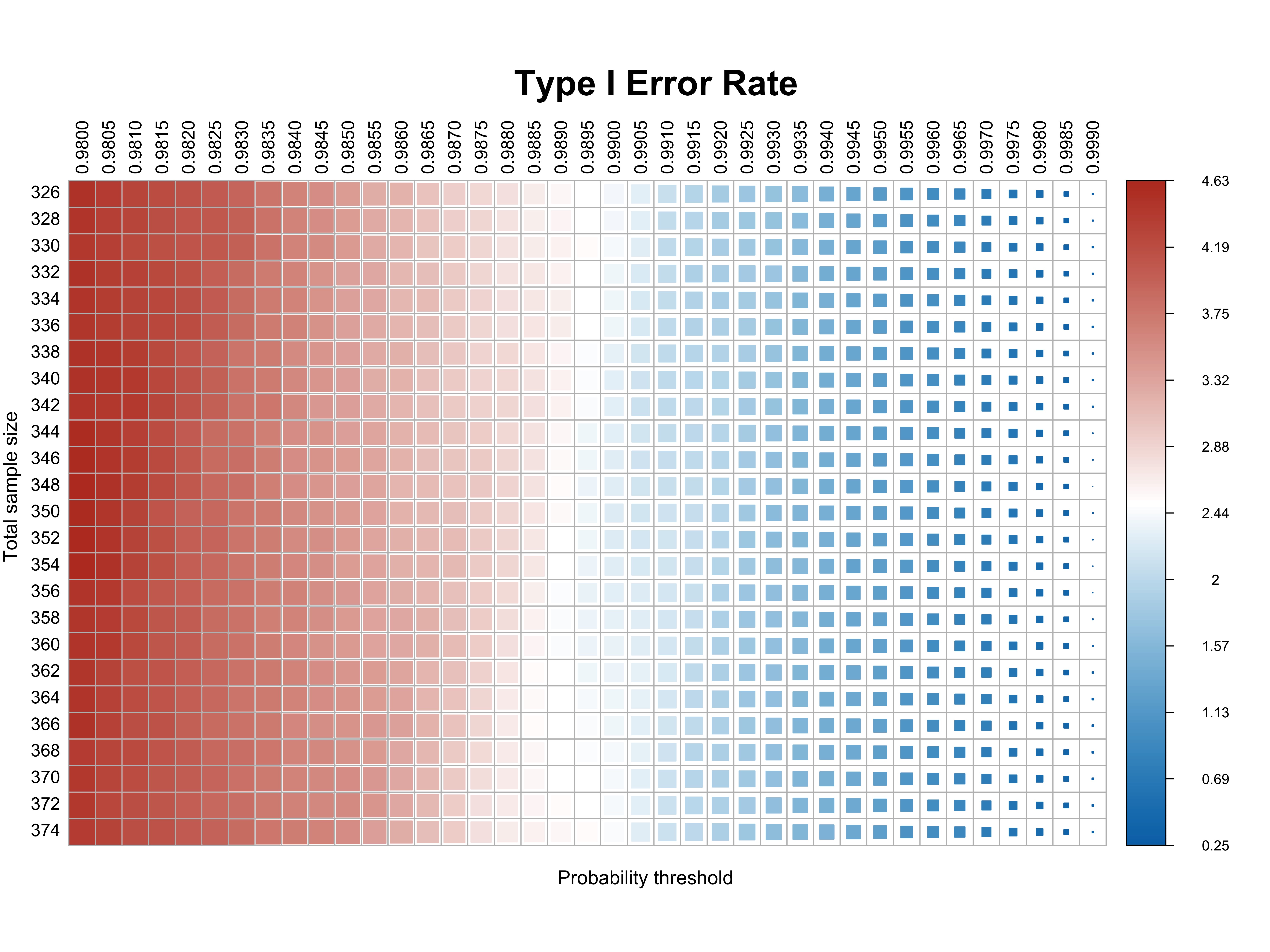}
	\caption{Overall type I error rates for combinations of posterior probability thresholds and total sample sizes in the conventional Bayesian GSD with two equally spaced interims.}
	\label{fig:21}
\end{figure}

\begin{figure}[p]
	\centering
	\includegraphics[width=\linewidth]{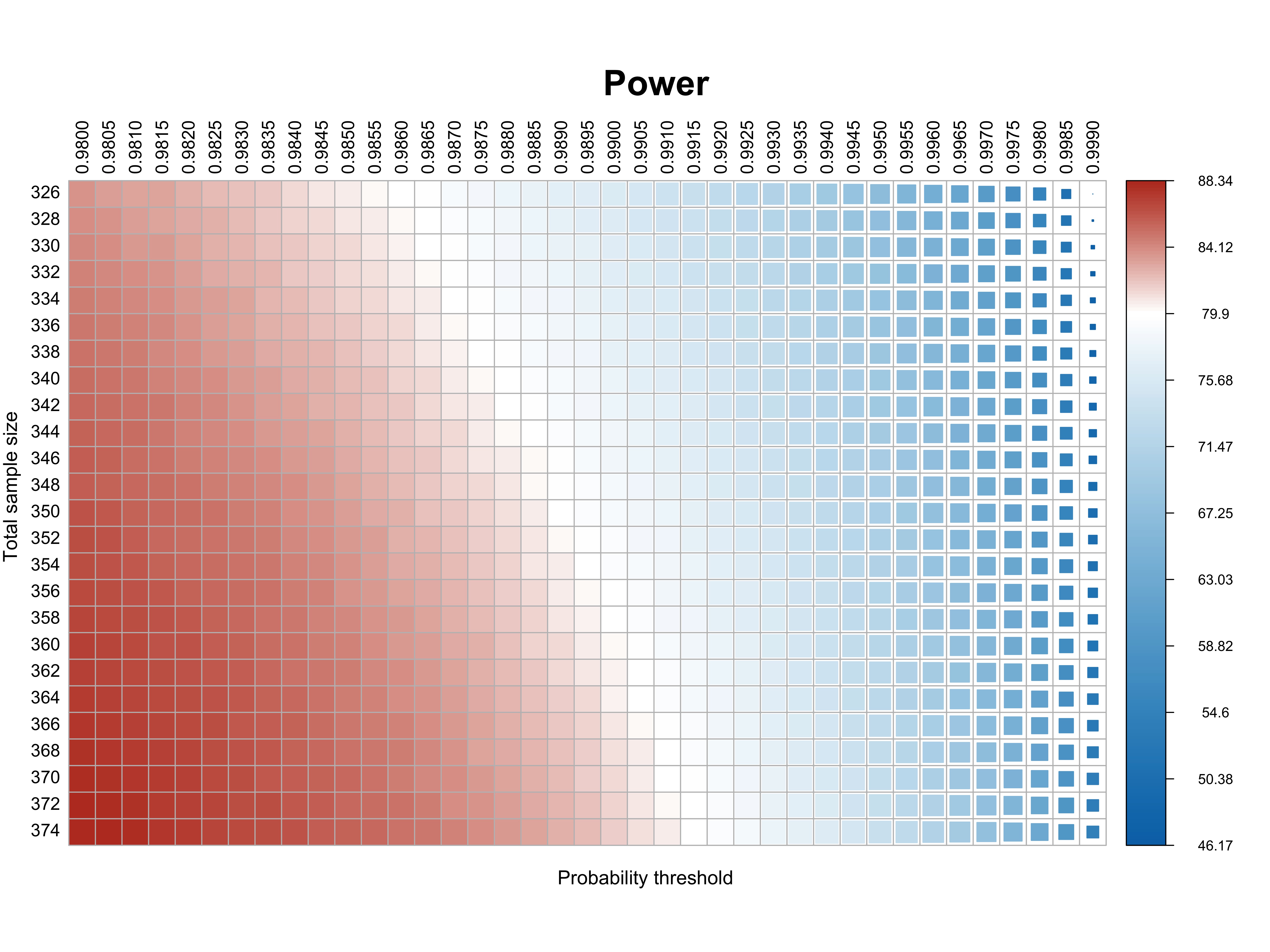}
	\caption{Powers for combinations of posterior probability thresholds and total sample sizes in the conventional Bayesian GSD with two equally spaced interims.}
	\label{fig:22}
\end{figure}

\begin{figure}[p]
	\centering
	\includegraphics[width=\linewidth]{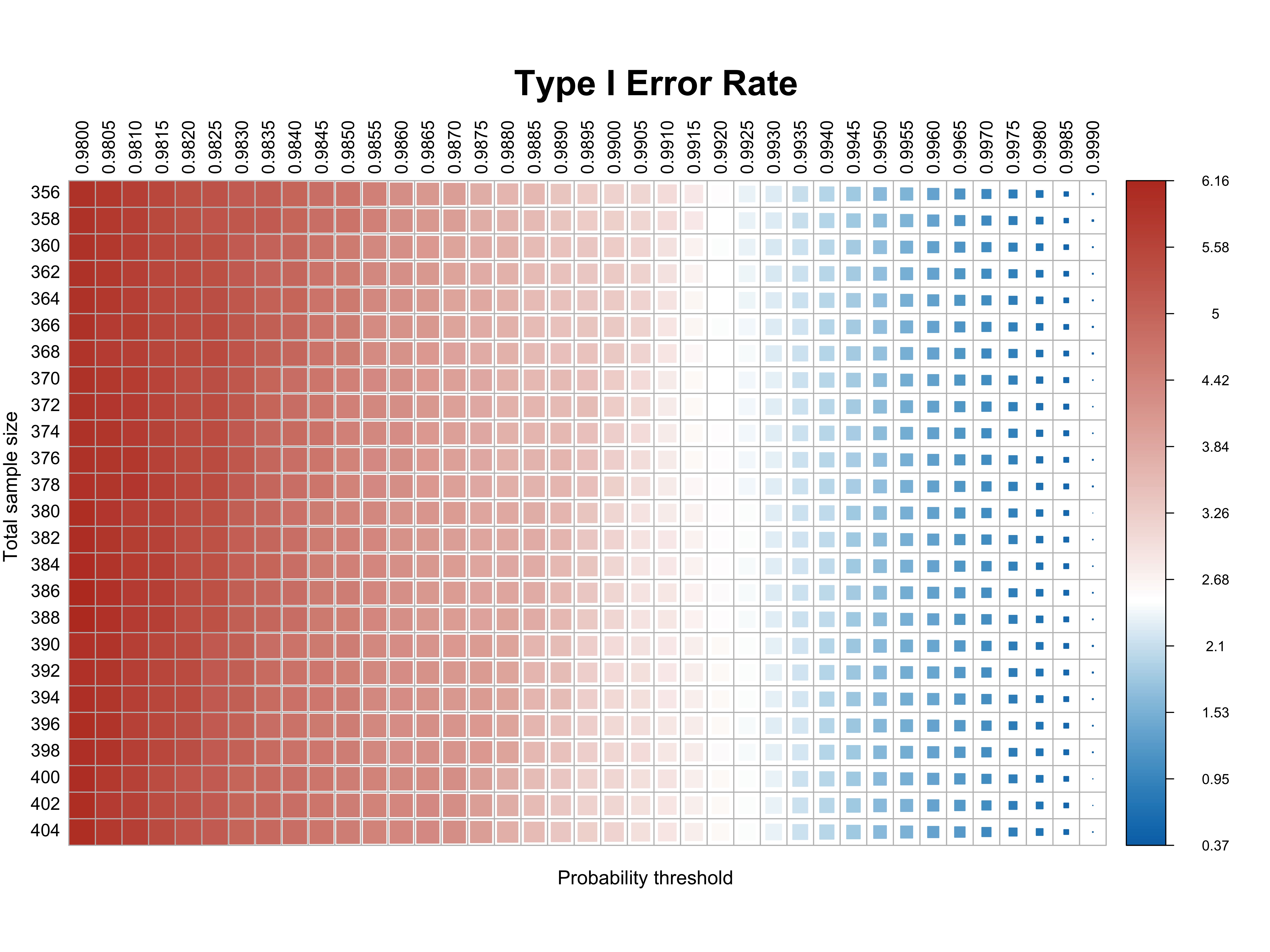}
	\caption{Overall type I error rates for combinations of posterior probability thresholds and total sample sizes in the conventional Bayesian GSD with four equally spaced interims.} 
	\label{fig:301}
\end{figure}

\begin{figure}[p]
	\centering
	\includegraphics[width=\linewidth]{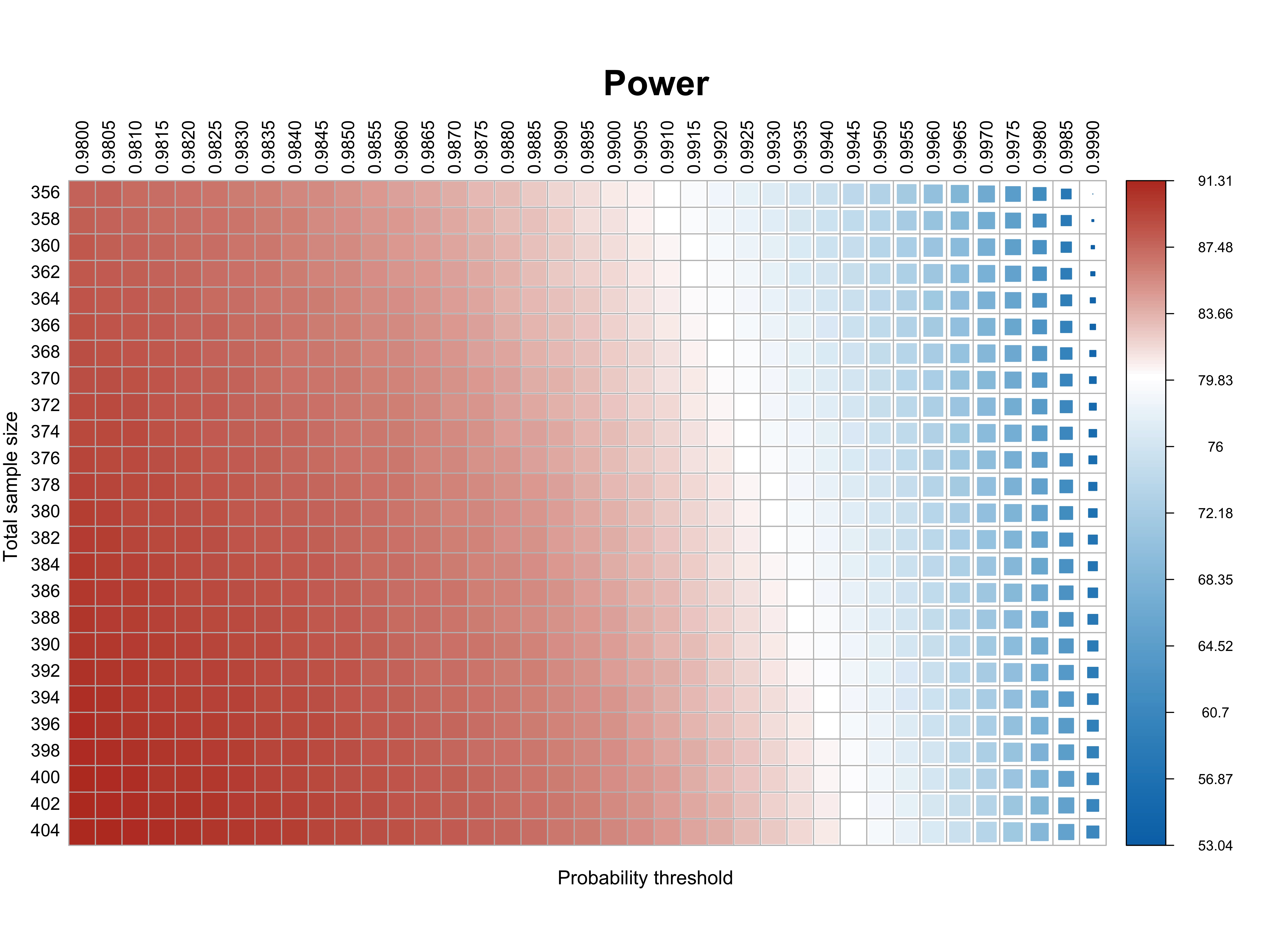}
	\caption{Powers for combinations of posterior probability thresholds and total sample sizes in the conventional Bayesian GSD with four equally spaced interims.}
	\label{fig:302}
\end{figure}

\begin{figure}[p]
	\centering
	\includegraphics[width=\linewidth]{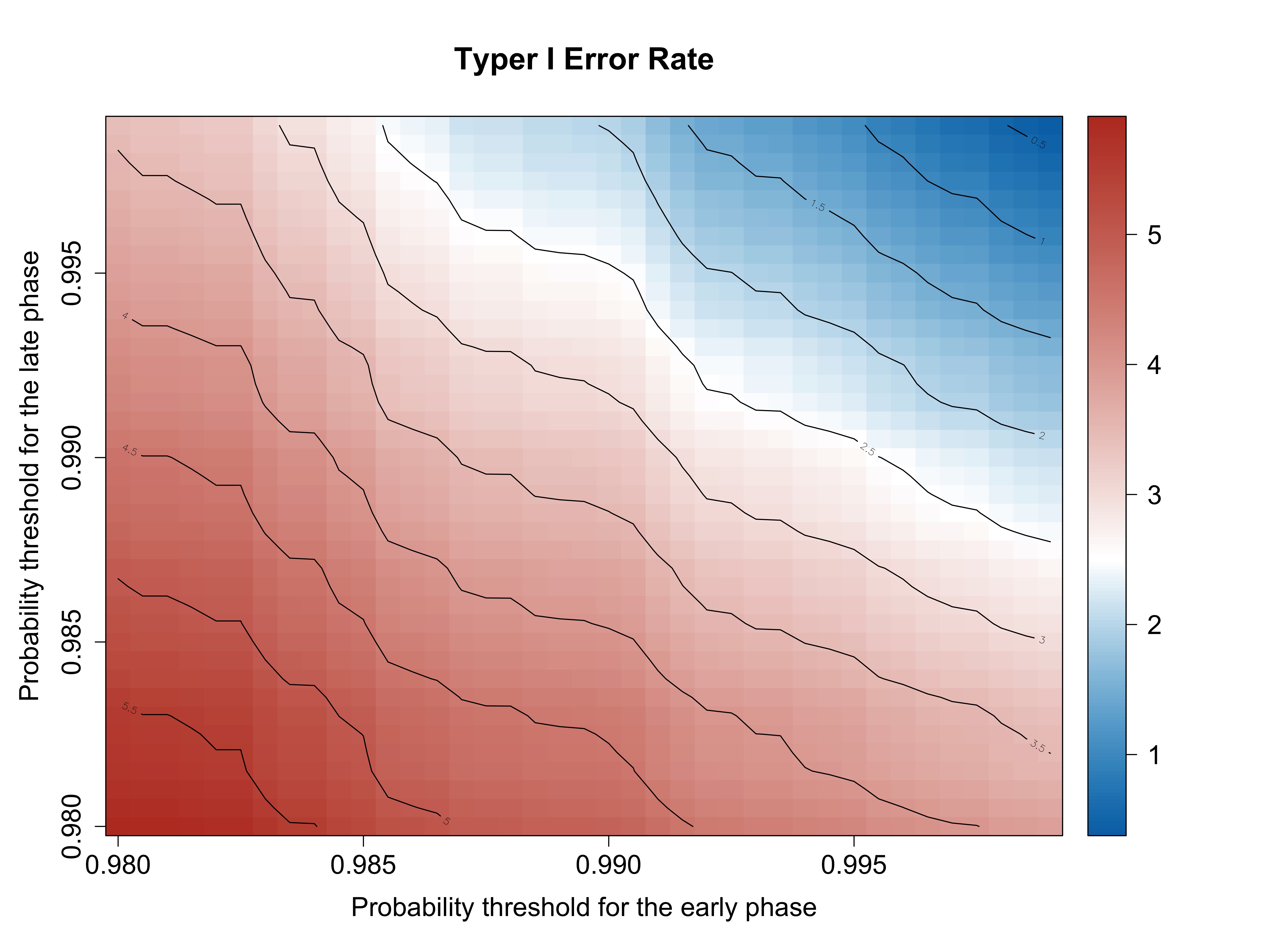}
	\caption{Overall type I error rates for combinations of early-phase and late-phase probability thresholds in the Bayesian GSD under Strategy 1.} 
	\label{fig:321}
\end{figure}

\begin{figure}[p]
	\centering
	\includegraphics[width=\linewidth]{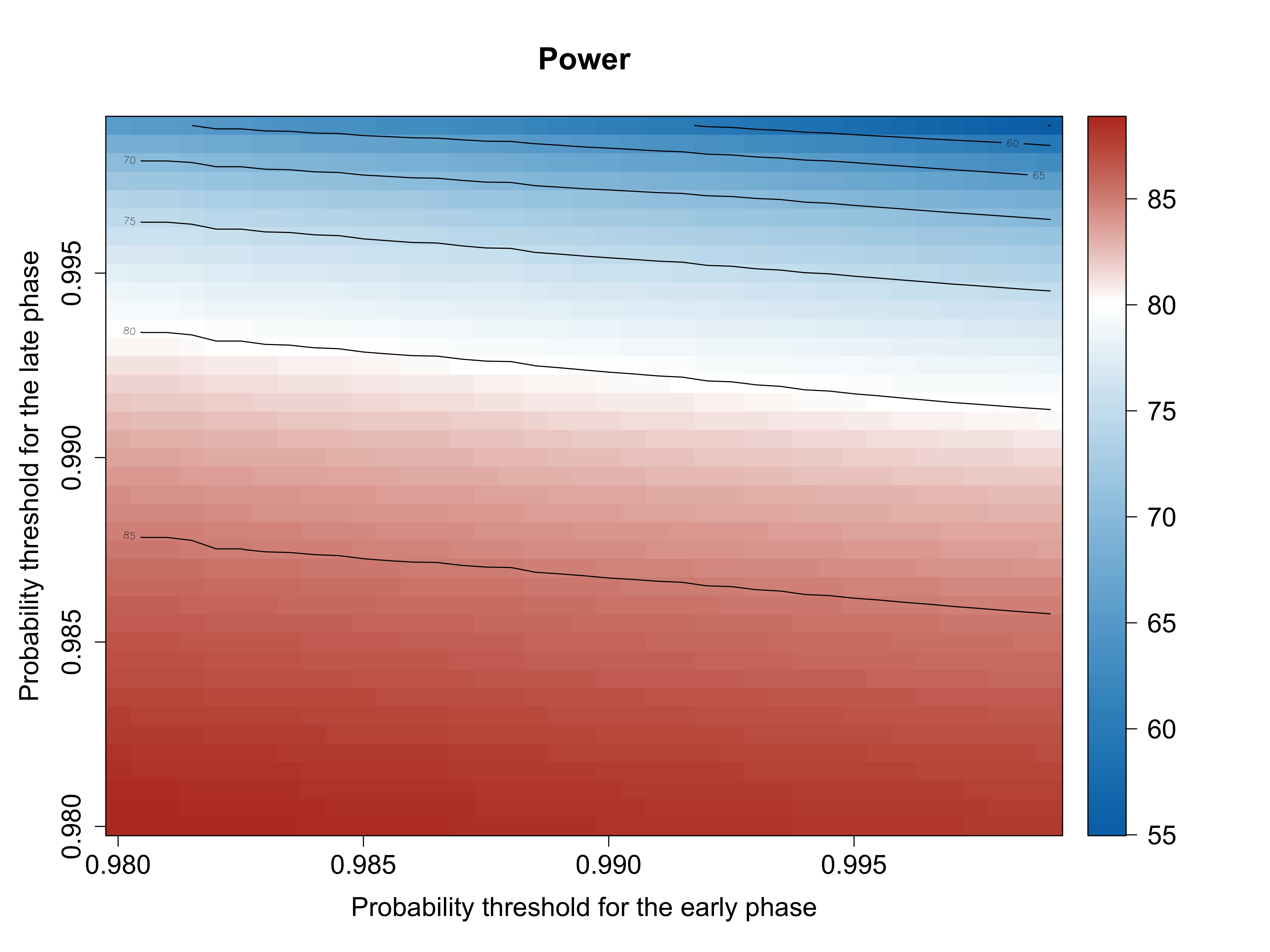}
	\caption{Powers for combinations of early-phase and late-phase probability thresholds in the Bayesian GSD under Strategy 1.}
	\label{fig:322}
\end{figure}

\begin{figure}[p]
	\centering
	\includegraphics[width=\linewidth]{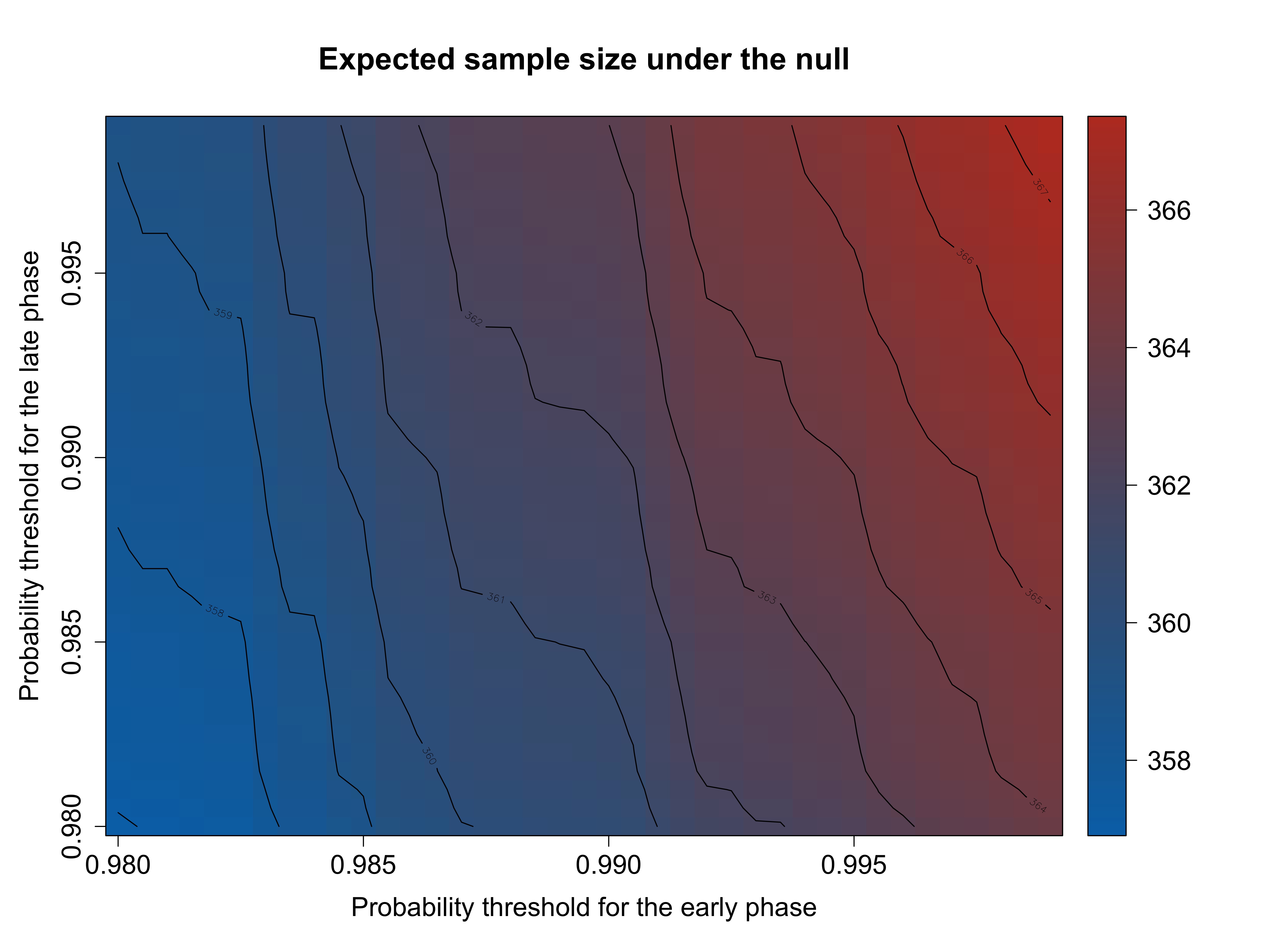}
	\caption{Expected sample sizes under the null for combinations of early-phase and late-phase probability thresholds in the Bayesian GSD under Strategy 1.}
	\label{fig:323}
\end{figure}

\begin{figure}[p]
	\centering
	\includegraphics[width=\linewidth]{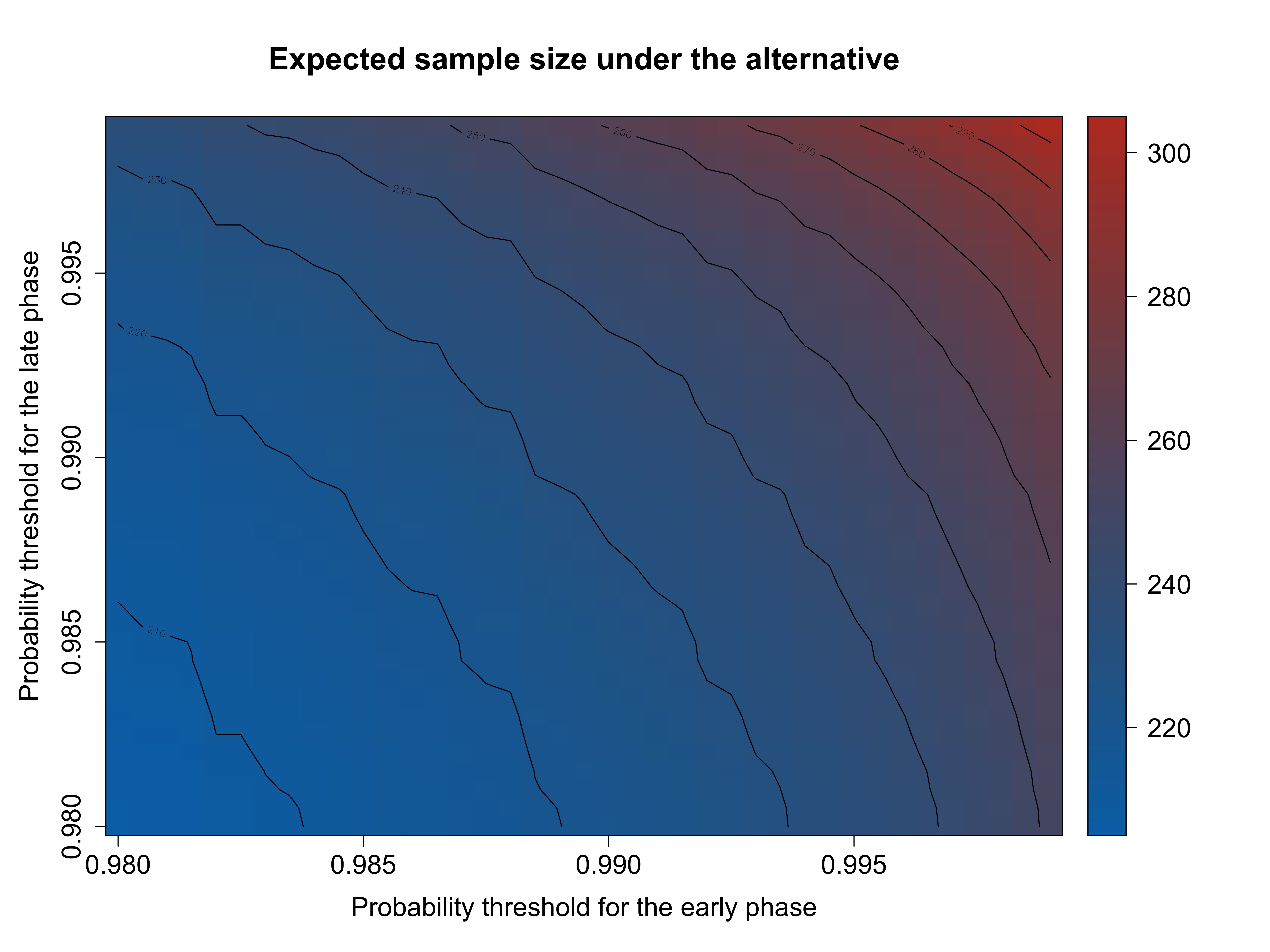}
	\caption{Expected sample sizes under the alternative for combinations of early-phase and late-phase probability thresholds in the Bayesian GSD under Strategy 1.}
	\label{fig:324}
\end{figure}

\begin{figure}[p]
	\centering
	\includegraphics[width=\linewidth]{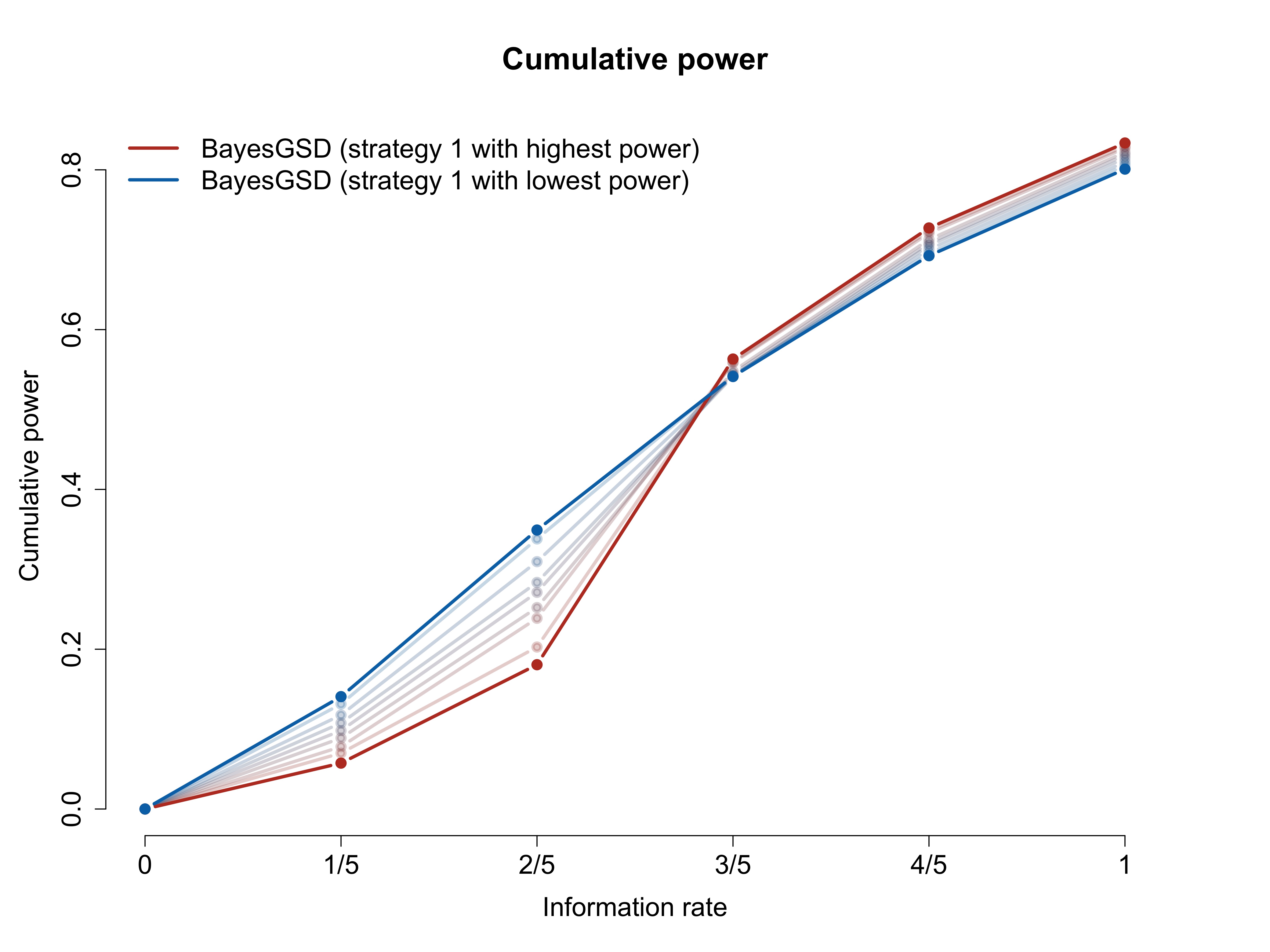}
	\caption{The cumulative beta-spending profiles for the calibrated Bayesian GSDs under Strategy 1. The Colour gradient reflects the associated power of each calibrated design: warmer (red) shades correspond to higher power, whereas cooler (blue) shades indicate lower power.}
	\label{fig:326}
\end{figure}

\begin{figure}[p]
	\centering
	\includegraphics[width=\linewidth]{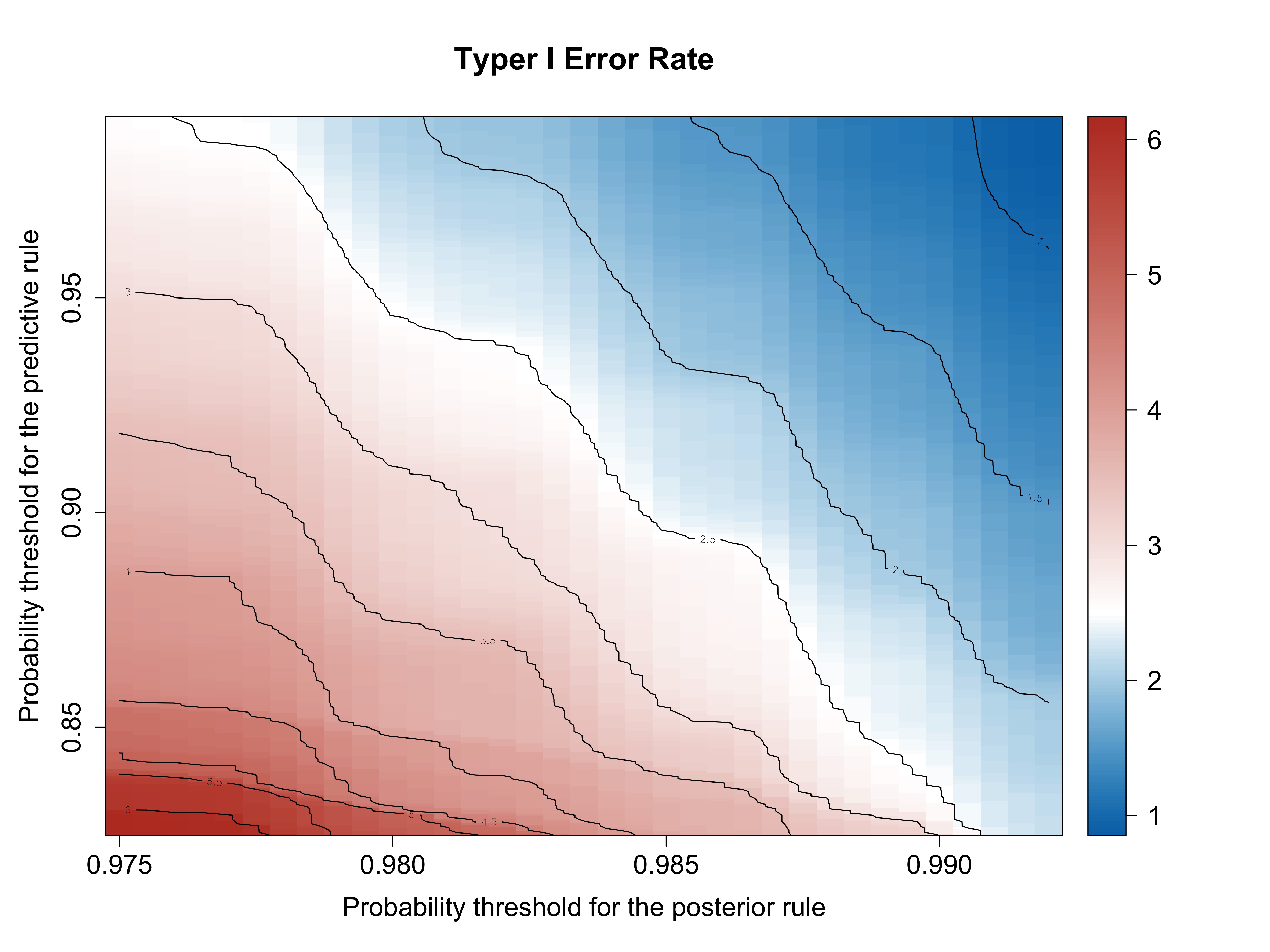}
	\caption{Overall type I error rates for combinations of early-phase and late-phase probability thresholds in the Bayesian GSD under Strategy 2.} 
	\label{fig:341}
\end{figure}

\begin{figure}[p]
	\centering
	\includegraphics[width=\linewidth]{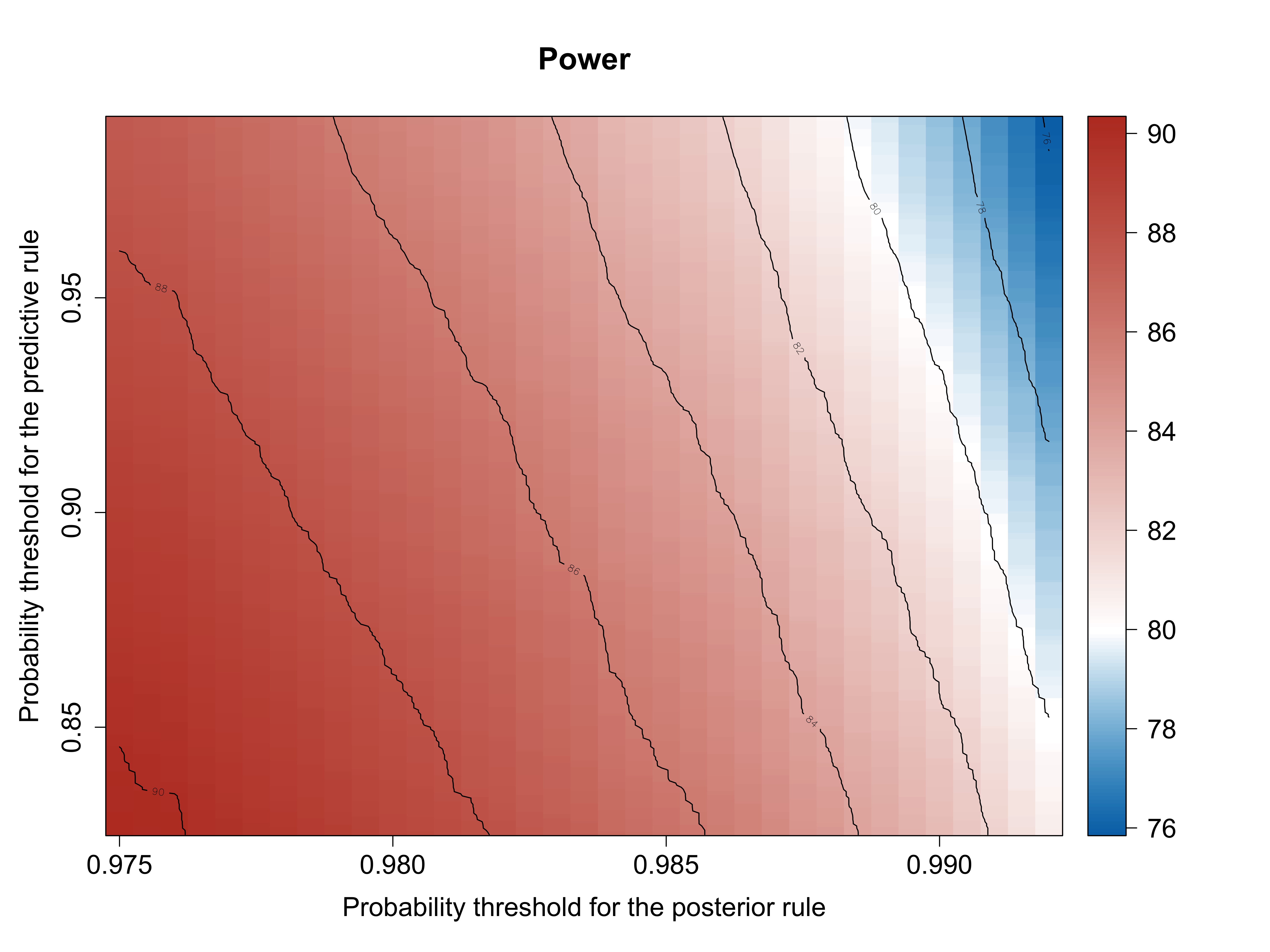}
	\caption{Powers for combinations of early-phase and late-phase probability thresholds in the Bayesian GSD under Strategy 2.}
	\label{fig:342}
\end{figure}

\begin{figure}[p]
	\centering
	\includegraphics[width=\linewidth]{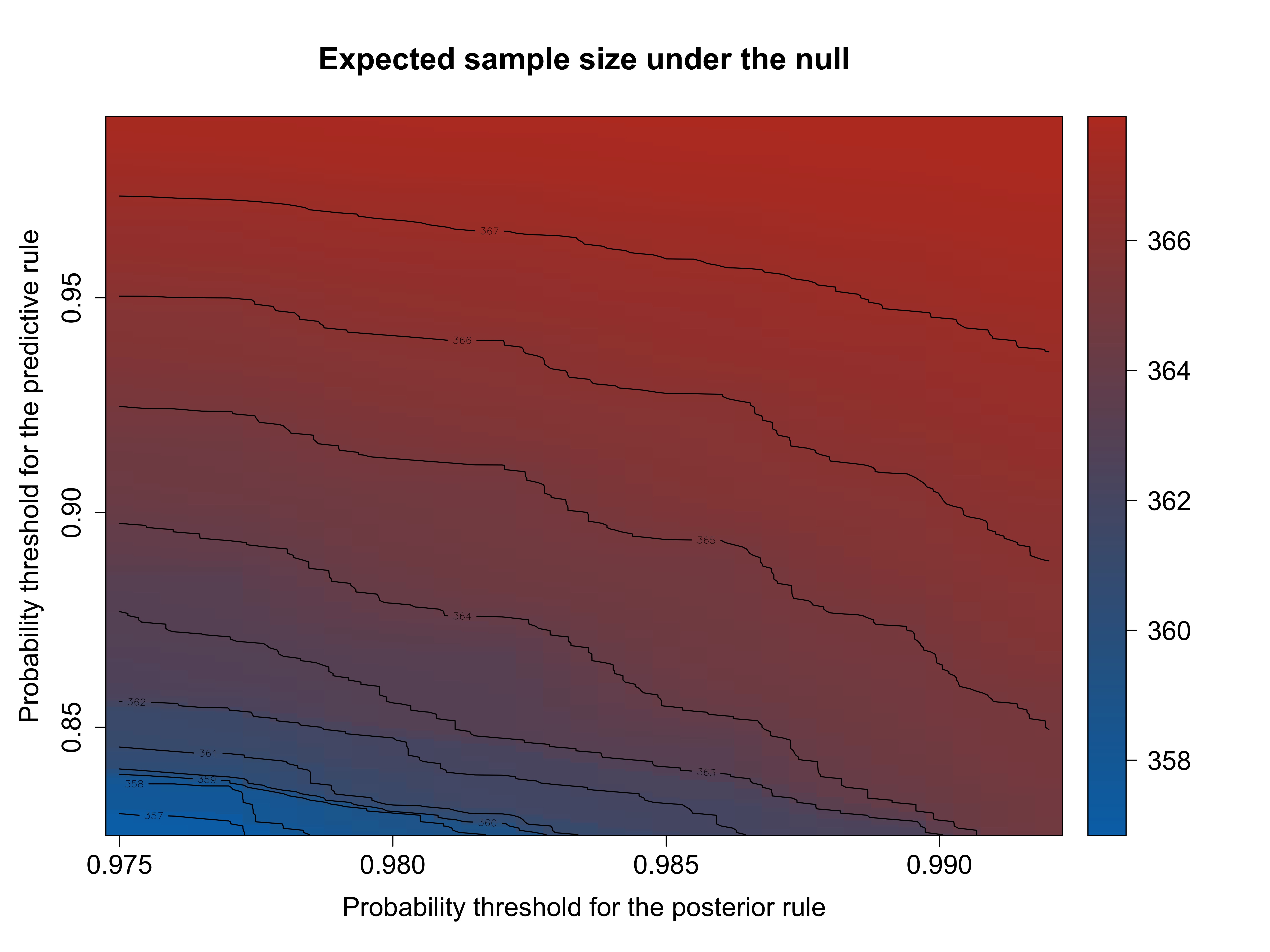}
	\caption{Expected sample sizes under the null for combinations of early-phase and late-phase probability thresholds in the Bayesian GSD under Strategy 2.}
	\label{fig:343}
\end{figure}

\begin{figure}[p]
	\centering
	\includegraphics[width=\linewidth]{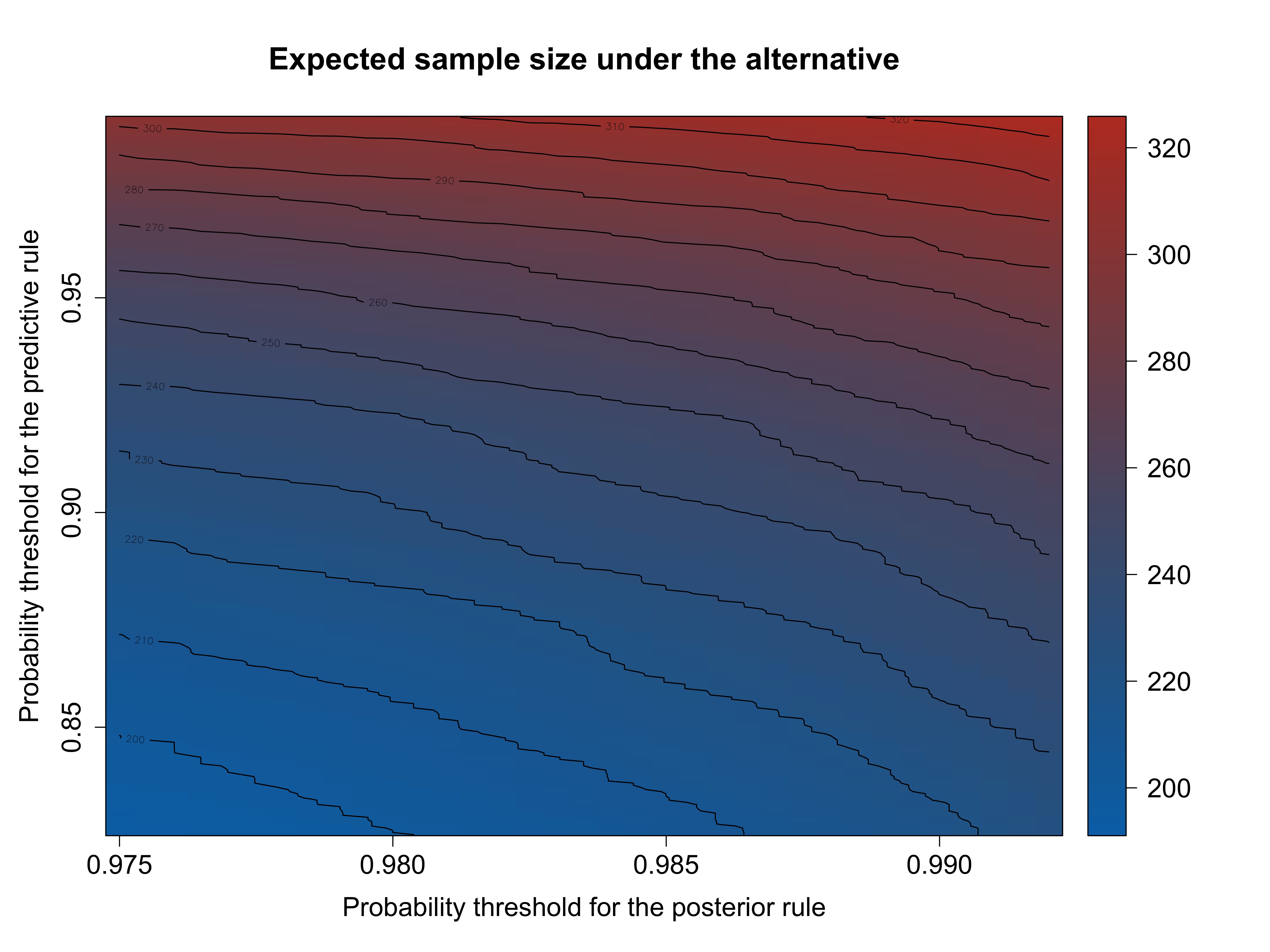}
	\caption{Expected sample sizes under the alternative for combinations of early-phase and late-phase probability thresholds in the Bayesian GSD under Strategy 2.}
	\label{fig:344}
\end{figure}

\begin{figure}[p]
	\centering
	\includegraphics[width=\linewidth]{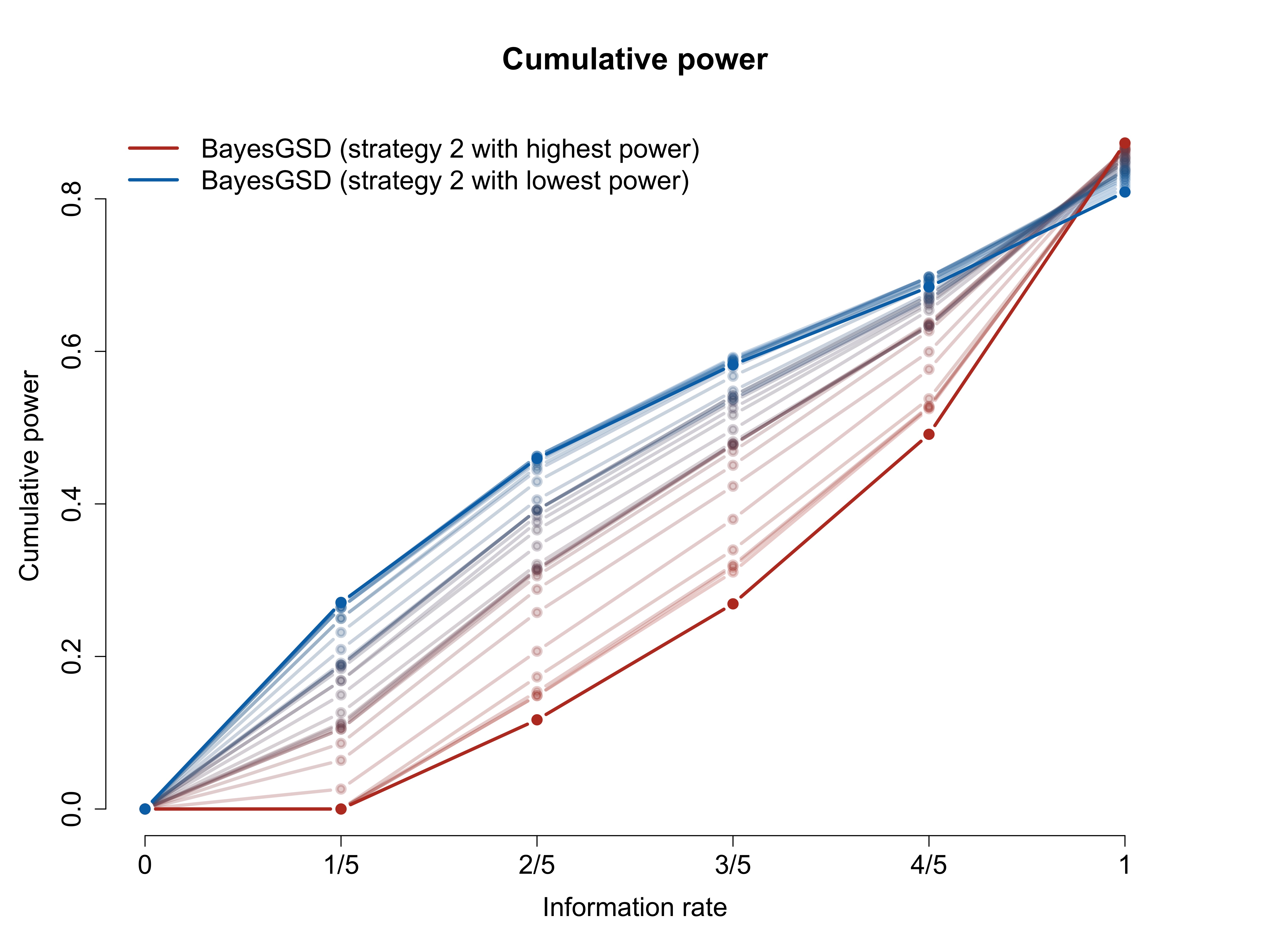}
	\caption{The cumulative beta-spending profiles for the calibrated Bayesian GSDs under Strategy 2. The Colour gradient reflects the associated power of each calibrated design: warmer (red) shades correspond to higher power, whereas cooler (blue) shades indicate lower power.}
	\label{fig:346}
\end{figure}

\end{document}